\documentstyle[prb,aps,twocolumn,amssymb]{revtex}
\begin{document}
\title{Full capacitance-matrix effects in driven Josephson-junction
arrays}
\author{Frank Gibbons$^1$, A. G\'ongora-T$^2$, and Jorge V. 
Jos\'{e}$^{1,2}$}
\address{$^{1}$Physics Department and Center for Interdisciplinary
Research on Complex Systems, \\ Northeastern University, Boston,
MA~02115, USA\cite{byline2}\\
$^{2}$Instituto de F\'{\i}sica,
Universidad Nacional Aut\' onoma de M\' exico,\\
Apdo. Postal 20-364, 01000~M\' exico D. F., M\' exico}
\date{\today}
\maketitle
\begin{abstract}
We study the dynamic response to external currents of periodic arrays
of Josephson junctions, in a resistively capacitively shunted junction
(RCSJ) model, including {\it full capacitance-matrix effects}. We
define and study three different models of the capacitance matrix
$C_{\vec{r},\vec{r}'}$: {\it Model A} includes only mutual
capacitances; {\it Model B} includes mutual and self capacitances,
leading to exponential screening of the electrostatic fields; {\it
Model C} includes a dense matrix $C_{\vec{r},\vec{r}'}$ that is
constructed approximately from superposition of an exact analytic
solution for the capacitance between two disks of finite radius and
thickness. In the latter case the electrostatic fields decay
algebraically. For comparison, we have also evaluated the full
capacitance matrix using the MIT {\sc fastcap} algorithm, good for
small lattices, as well as a corresponding continuum effective-medium
analytic evaluation of a finite-voltage disk inside a zero-potential
plane. In all cases the effective $C_{\vec{r},\vec{r}'}$ decays
algebraically with distance, with different powers. We have then
calculated current--voltage characteristics for DC+AC currents for all
models. We find that there are novel giant capacitive fractional steps
in the I-V's for {\it Models B} and {\it C}, strongly dependent on the
amount of screening involved. We find that these fractional steps are
quantized in units inversely proportional to the lattice sizes and
depend on the properties of $C_{\vec{r},\vec{r}'}$. We also show that
the capacitive steps are not related to vortex oscillations but to
localized screened phase-locking of a few rows in the lattice.  The
possible experimental relevance of these results is also discussed.
\end{abstract}


\section{Introduction}

\label{sec:intro} 
There has been considerable recent interest in the study of the
dynamic response of two-dimensional Josephson-junction arrays (JJA)
under different physical conditions. This interest has both
theoretical and experimental motivations\cite{rev}. Experimentally,
recent advances in photolithographic micro-fabrication techniques have
allowed the manufacture of these arrays with specific tailor-made
properties\cite{rev}. The arrays can be made with proximity-effect,
i.e., superconducting\---metal\---superconducting (SNS) junctions, or with
superconducting\---insulating--superconducting (SIS) junctions. In the
SNS case there is essentially zero capacitance between the
superconducting islands. In the SIS case the electrodes that form the
junctions have nonzero self and mutual capacitances that can, as
discussed in this paper, significantly influence the physical
properties of the arrays. Theoretically, the arrays are described in
terms of a large set of coupled, non-linear differential equations,
overdamped in the SNS case and underdamped in the SIS case. The SNS
arrays have been studied more extensively, for they can be fabricated
with more ease, better uniformity and larger lattice sizes. In the SNS
arrays the critical currents can be large and thus self-induced
magnetic fields must be taken into account, through the inclusion of
Faraday's induction law to fully describe the experimental
systems\cite{nos4,phillips,lachenmann}. In contrast, in SIS arrays the
critical currents are relatively low and self-induced fields are
usually not significant. In this paper we consider the SIS case where
the electrostatic properties of the junctions in the array are of
importance, while the inductances are not included here.

When the arrays are driven by DC+AC currents, non-equilibrium
stationary coherent oscillatory vortex states may appear, both in the
SNS and SIS arrays. Experimentally, giant Shapiro steps (GSS) have
been observed in the I-V characteristics of SNS arrays in
zero\cite{clark} and rational magnetic field
frustrations\cite{benz}. The frustration is defined by
$f=\Phi/\Phi_0$, where $\Phi$ is the average applied magnetic flux per
plaquette, and $\Phi_0=h/2e$ is the magnetic flux quantum. Coherent
oscillations of ground-state field-induced vortices are responsible
for the existence of the {\it fractional} GSS when $f=p/q$, with $p$
and $q$ relative primes\cite{benz}. This interpretation was
successfully verified in numerical simulations when the junctions in
the array were modeled by the resistively shunted junction (RSJ)
model\cite{numshap1,numshap2}. Half-integer steps have also been found
in zero-field topologically disordered SNS arrays, due to a special
oscillatory vortex pattern termed the {\it axisymmetric coherent
vortex state}, or ACVS\cite{acvs}. Fundamentally different
half-integer Shapiro steps have also been seen in capacitive array
models, with $f=0$, when the capacitance matrix includes only the
mutual capacitance between superconducting islands\cite{hagenaars}. In
all the cases discussed above the steps found in the I-V
characteristics were microscopically due to the collective coherent
oscillatory motion of dynamically or magnetically induced vortices.

Experimental papers on SIS arrays do report on the importance of
electric field screening, or capacitive effects, in particular when
the junctions are of submicron size\cite{mooij}, but also when they
are large\cite{benz2}. In the cases when the electrostatic screening
length is smaller than the array size, the screening may affect the
array's transport properties in a significant way. Most work that has
included capacitive effects, however, has done so considering only
nearest-neighbor mutual capacitances. To be more specific, consider an
array formed by $N$ conducting islands where the capacitance of the
whole array can be characterized by a matrix $C_{\vec{r},\vec{r}'}$,
where $\vec{r}$ and $\vec{r}'$ are vectors denoting the locations of
two electrostatically interacting islands. The dynamical equations of
motion for a driven array that includes general capacitive affects can
be written as an extension of the resistively capacitively shunted
junction (RCSJ) model. This model, which includes mutual capacitances,
has been successful in explaining the giant Shapiro steps described
above. The corresponding equations read:
\begin{eqnarray}
\beta_c\sum_{\vec{r}'}C_{\vec{r},\vec{r}'}{\ddot{\theta}_{\vec{r}'}} 
&+ &\sum_{\vec{r}'}G_{\vec{r},\vec{r}'}^{-1}\dot{\theta}_{\vec{r}'}
+ \sum_{\hat{\mu}}\sin \left( \theta_{\vec{r}+\hat{\mu}}
- \theta _{\vec{r}}\right)\nonumber\\
& =& 
i_{\text{ext}}(\vec{r},t) \equiv i_{dc} + i_{ac}\sin (2\pi \nu t).
\label{eqn(1)}
\end{eqnarray}
Here $\theta _{\vec{r}}$ is the phase of the Ginsburg--Landau order
parameter for the ${\vec{r}}$-th island; $G^{-1}$ is the inverse
lattice Green function in two dimensions (2D) (i.e., the discrete
Laplacian); the $\sum_{\hat{\mu}}$ indicates a sum over nearest
neighbors. $i_{\text{ext}}(\vec{r},t)$ is the external current
injected at each lattice site $\vec{r}$, with both a DC and an AC
component that oscillates at frequency $\nu $. The currents are
expressed in units of the junction critical current $I_c$; time is
measured in units of the characteristic time
$1/\omega_c=\hbar/(2eRI_c)$, with $R$ the normal-state junction
resistance, $e$ the electronic charge, and $\hbar =h/{2\pi}$, with $h$
Planck's constant. The capacitance matrix entries are normalized by
the single-junction dissipation, or Stewart--McCumber, parameter
$\beta _c={2eR^2I_cC_m}/\hbar$, where $C_m$ is the mutual capacitance
of a single junction.

The purpose of this paper is to study solutions to Eq.~(\ref{eqn(1)})
for different ever-more-realistic approximations to the
$C_{\vec{r},\vec{r}'}$ matrix. Just as was done in the inductive
case\cite{nos4,phillips,modb}, here we consider the array response to
external currents when $C_{\vec{r},\vec{r}'}$ goes from being short-
to long-ranged. We concentrate mostly on the r\'egime that does not
show chaotic solutions in the single-junction case, for extra
complications may arise in that case that can complicate the analysis
further. We also only consider the semiclassical regime, where the
Josephson energy dominates the charging energy. As in previous studies
we do find giant steps in the I-V characteristics in the models
considered here. In contrast, however, to the GSS described above, the
ones described in this paper are not associated with collective vortex
oscillations. Instead, depending on the specific model for the
capacitance matrix, they are related to the phase-locking of specific
row of junctions to the external drive.

Here we introduce and study different approximate models for the
capacitance matrix. We know that we cannot exactly calculate the
general $C_{\vec{r},\vec{r}'}$ matrix by simply giving the geometric
configurations of the conductors. We start then by assuming that the
$C_{\vec{r},\vec{r}'}$ has only the nearest-neighbor mutual component,
which reduces $C_{\vec{r},\vec{r}'}$ to a tridiagonal form. This is
the model that has been studied most in the past (see
{\bf Ref}.~\onlinecite{hagenaars}). Including the self-capacitances, which
can be done experimentally by putting a ground plane underneath the
array, leads to screening in $C_{\vec{r},\vec{r}'}$. The self- plus
mutual-capacitance matrix approximation reduces the complexity of the
full $C_{\vec{r},\vec{r}'}$ significantly. Making this approximation
physically means that the electric field lines between charges will be
confined to the plane of the array, with logarithmic interaction with
distance.  Here we want to go beyond these approximations and take
into account the fact that the field lines are experimentally
three-dimensional in nature, so we need to consider further elements
in $C_{\vec{r},\vec{r}'}$.  We must then resort to approximate
representations of this matrix.  We have followed several different
ways to estimate the behavior of $C_{\vec{r},\vec{r}'}$, some
analytic, some numeric. On the analytic side, we evaluated the
capacitance of two disks of finite thickness and radius in a
plane\cite{jackson,ley}. We use the result obtained from this
calculation to fill in, using approximate superposition, what would be
the matrix elements of a full $C_{\vec{r},\vec{r}'}$ matrix.  This
approach is similar to the one followed in the full inductance
calculations that gave rather good results\cite{modb}. To ascertain
the nature of this approximation, next we used the MIT {\sc fastcap}
algorithm\cite{fastcap} that numerically evaluates the
$C_{\vec{r},\vec{r}'}$ for small systems. To further understand the
{\sc fastcap} results we also evaluated the capacitance of a finite
potential disk of finite radius embedded in an infinite grounded
plane.  We looked at this problem since it represents a type of
effective-medium approximation of the {\sc fastcap} algorithm. In all
these estimates of the $C_{\vec{r},\vec{r}'}$ matrix we found that it
decays algebraically with distance. The rate of decay, i.e., the power
of the decay is different in the various model approximations.  This
leads to some quantitative difference in the corresponding I-V
results, but we expect that qualitatively they are correct, and
certainly when compared with the exponentially decaying behavior of
the self--mutual approximation to $C_{\vec{r},\vec{r}'}$.

In studying the behavior of the I-V's of the full-capacitance model,
we have found other types of giant capacitive voltage steps in the I-V
curves, which are of a very different origin from the ones studied
previously. In this paper, we define a {\em row} as a series of
neighboring junctions along the direction transverse to the
current. Under certain conditions, we have observed these {\it giant
capacitive fractional steps} (GCFS), which are caused by {\it
phase-locking} of junctions in a given row to the external drive, with
no significant phase-locking between rows. The array exhibits the
phase-locking behavior which underlies the GCFS, but we find that this
does not occur throughout the entire array. Rather, it only occurs in
certain rows near the edge of the array, due to screening. We discuss
these results in more detail in the main body of the paper.

The outline of the paper is the following: In Section~II we present
the equations that describe the different models mentioned above, as
well as briefly discussing the methods we used to study them. We also
present some analytical results which we used in our analysis and to
fill out the entries of the full capacitance matrix. Section~III
discusses the bulk of our results and their analysis, for the cases
where the driving current is purely DC or a combination DC+AC. We
study the I-V's, spectral functions, and topological phase and current
distributions, to support our view that the steps are indeed generated
by phase-locking of a few rows of junctions (rather than the entire
array), and not by the presence of a stable vortex configuration, as
is the case with other giant fractional steps. Finally, in Section~IV
we summarize and discuss our results.

\section{Capacitance--matrix models}

\label{sec:eqnmethod}
In this section we proceed to define the different capacitance matrix
models studied and compared in this paper. Our goal is to solve
Eq.~(\ref{eqn(1)}) for different $C_{\vec{r},\vec{r}'}$ matrix
models. We begin by converting the set of $N\equiv L_x\times L_y$
second-order ordinary differential equations into $2N$ first-order
equations. Explicitly, we define the variables $z_{\vec{r}}$ and
$\dot{z}_{\vec{r}}$ that satisfy the equations,
\begin{eqnarray}
z_{\vec{r}} & = &\dot{\theta}_{\vec{r}},  \label{eq:firstorderA}\\
\dot{z}_{\vec{r}}
&=&\frac{1}{\beta_c} \sum_{\vec{r}'} C_{\vec{r},\vec{r}'}^{-1} \Bigg(
i_{\text{ext}}(\vec{r}')-\sum_{\vec{r}''}G_{\vec{r}',\vec{r}''}^{-1}
z_{\vec{r}''} \nonumber\\
&&\qquad\qquad\qquad\qquad -  \sum_{\hat\mu} \sin \left( \theta_{\vec{r}' + \hat\mu} -
\theta_{\vec{r}'} \right) \Bigg),
\label{eq:firstorderB}
\end{eqnarray}
where the indices $\vec{r},\vec{r}'$, and $\vec{r}''$ run from $1$ to
$N$, that is, over the whole array. Periodic boundary conditions are
used in the direction perpendicular to the current. A schematic of the
arrays under study is shown in Fig.~\ref{array_fig}. In order to
simulate large arrays, it is useful to use the general properties of
the capacitance matrix that arise from the positivity of the total
electrostatic energy of the array of electrodes. Specifically,
$C_{\vec{r},\vec{r}'} > 0, \forall \ \vec{r}=\vec{r}'$,
\mbox{$C_{\vec{r},\vec{r}'} < 0$}, $\ \forall \ \vec{r}\neq \vec{r}'$,
and $C_{\vec{r}, \vec{r}}\ge \sum_{\vec{r}'\neq
\vec{r}}|C_{\vec{r},\vec{r}'}|$. In addition to these properties, we
have translational invariance in a periodic array, so that we can
write $C(\vec{r},\vec{r}')\equiv C(|\vec{r}-\vec{r}'|)$, which allows
for a significant simplification in the calculation of
$C_{\vec{r},\vec{r}'}$. Our approach parallels the scheme used in the
dynamic study of inductive Josephson-junction arrays\cite{nos4}, and
we consider here the following three distinct capacitive models:

{\it Model A}: In this case we consider only the contributions by the
mutual capacitances between electrodes in the array. Here the
capacitance matrix is simply tridiagonal, i.e.,
\[ 
C_{\vec{r},\vec{r}'}=4\delta _{\vec{r},\vec{r}'}
-\delta_{\vec{r},\vec{r}'\pm \hat{e}_x}
-\delta _{\vec{r},\vec{r}'\pm \hat{e}_y}
\equiv G_{\vec{r},\vec{r}'}^{-1}
\]
with $\hat{e}_x$ and $\hat{e}_y$ are unitary vectors along the $x$-
and $y$-directions. This model has been studied by many
authors\cite{rev,hagenaars,modb}, both for ordered and disordered
arrays, and some results are already known.

{\it Model B}: This model extends {\it Model A} by including the
self-capacitance term in the capacitance matrix as:
\[ 
C_{\vec{r},\vec{r}'}=(4+r_c)\delta _{\vec{r},\vec{r}
'}-\delta _{\vec{r},\vec{r}'\pm \hat{e}_x}-\delta _{\vec{r},\vec{
r}'\pm \hat{e}_y},
\]
where $r_c=|C_s/C_m|$, and $C_s$ and $C_m$ are the self-capacitance
and the mutual capacitances, respectively. This model tends to {\it
Model A} in the limit $r_c\rightarrow 0$. In the continuum limit the
inverse of the capacitance matrix becomes
\[
C^{-1}(r)\simeq \frac{1}{2\pi C_m}\, K_0(\sqrt{r_c}\, r),
\]
with $K_0(z)$ is the modified Bessel function of zeroth order.
Asymptotically, we have that $K_0(z)\sim -\ell n z$, for $z\ll 1$, and
$K_0(z)\sim \sqrt{\frac{\pi}{2z}}e^{-z}$, for $z\gg 1$. We see that
when the self-capacitance is very small the interactions are
essentially logarithmic with distance, and for large $C_s$ they are
exponential.  We define then the screening length as $\Lambda
_0=\frac{1}{\sqrt{r_c}}$. We will describe our results for the models
studied here as functions of the parameters, $\beta_c$, $\nu$ and
$\Lambda_0$, or $r_c$.

{\it Model C}: Here we consider the capacitive effects of all
conductors on each other. This means that the capacitance matrix is
dense, and the specific form of the individual entries depends on the
details of how we model the conductors in the array. We have used
several routes to analyze the long-range interaction that gives
approximate full capacitance-matrix models. Most of our I-V
calculations are based on a $C_{\vec{r},\vec{r}'}$ obtained by an
approximate superposition model from an exact analytic solution for
the capacitance of two circular disks of finite thickness and
radius\cite{jackson,ley}. We review the derivation of this result in
detail in the Appendix.

We place each disk at a lattice site in a square array. For the total
capacitance matrix we use simple superposition of the capacitances
between any two disks, as done in the inductive case, so as to get
\begin{equation}
C_{\vec{r},\vec{r}^{\prime}}=C(i,j) = \frac{1}{4\tanh^{-1}\left( \sqrt{1 - 
\frac{D^2}{i^2+j^2}} \right) },\label{2cap_disk1}
\end{equation}
and
\begin{equation}
C_m \equiv C(i=1, j=0) = \frac{1}{4\tanh^{-1}\left( \sqrt{1 - D^{2}}\right)},
\label{2cap_disk2}
\end{equation}
where $\vec{r}- \vec{r}' = i\hat e_x + j\hat e_y$, $i$ and $j$ are
integers, $D$ is the diameter of a disk and $R$ is the distance
between their centers (see Fig.~\ref{packing}). Note that the
magnitude of $D$ defines the packing of the disks as shown in the
figure. The two-disk expression for $C(i,j)$ given above allows us to
go from weak to strong screening by just varying $D$. For different
values of $D$, $C(i,j)$ has an initial rapid decay with distance and
then it decays algebraically. We show in Fig.~\ref{caps_power} the
behavior of the effective full $C(i,j)$ matrix for some values of $D$.

For comparison to the two-disk model result, we also have used the
{\sc fastcap} numerical capacitance extraction tool developed at
MIT\cite{fastcap}. This algorithm quantitatively calculates the
capacitance matrix for a given distribution of conductors. Whan et al.
[\onlinecite{whan}] have studied one and two-dimensional arrays
of circular dots using {\sc fastcap}. They also obtained a power-law decay
with exponent close to one. We also use their result in the
evaluation of DC IV's in Section III. We have also explicitly used {\sc
fastcap}, to compute $C_{\vec{r},\vec{r}'}$ for an $8\times8$
array of square plates of various thicknesses, ranging from 1000~\AA{}
to 35000~\AA. We took the dimensions used by van der Zant {\em et
al.}\cite{vanderZant:thesis}. A log--log plot of the capacitance
versus the distance is shown in Fig.~\ref{caps_power}. This
capacitance result tends to the two-disk array model for a specific
value of $D$, but differs for others. The essence of the result,
however, is that the off-diagonal capacitance matrix decays
algebraically.

The {\sc fastcap} approach is limited in that the arrays that can be
simulated are relatively small. (For example, $11\times 11$ was the
largest size possible on a fast DEC Alpha with 196~MB RAM.)  In the
{\sc fastcap} algorithm the capacitance matrix entries are calculated
iteratively, by successively grounding all but one of the conductors,
and holding it at a fixed nonzero potential.  The limitations of this
approach are that it is time-consuming and not many parameter values
can be easily considered. In contrast, the two-disk model is more
flexible and allows the study of larger lattices and a wider range of
parameter values.

In order to further understand how the capacitance matrix behaves we
have also analytically calculated the electrostatic potential, and
then the capacitance, of a circular disk of finite width set at a
fixed potential embedded in a conducting plane set at zero potential
The disk is of radius $s=D/2$ located at the coordinate origin and at
a finite nonzero potential $V_0$, while the conducting plane is at
zero potential. We considered this problem since it is, in a sense,
similar to the lattice algorithm used in {\sc fastcap} and it also
gives us another evaluation of the capacitance in the continuum
limit. We want to calculate the potential outside of the plane and
thus start from the general expression for the potential at point
$\vec{r}$ given by (see Jackson\cite{jackson})

\[
V(\vec{r})=-\frac {1}{4\pi }\oint_S V(\vec{r}')\frac{\partial 
F(\vec{r},\vec{r}')}{\partial z'}{\Big|_{\vec{n} =\vec{0}}}\,\,dA'. 
\]
Here $F(\vec r,\vec{r}')$ is the Green function for two unit charges
at $z$ and $z'$, ${\vec n}$ is the normal vector to the surface of
integration $S$, and $dA'$ the surface differential. In cylindrical
coordinates $(\rho ,\varphi ,z)$, $F(\vec{r},\vec{r}')$ is given by

\begin{eqnarray*}
F(\rho, \rho', \phi, \phi', z, z') &= &
\frac {1}{\sqrt{\rho^2 + \rho'^2-2\rho \rho'\cos \gamma +(z-z')^2}}\nonumber\\
& & -\frac {1}{\sqrt{\rho^2 + \rho'^2 - 2\rho\rho'\cos\gamma
+(z+z')^2}},
\end{eqnarray*}
where $\gamma =(\varphi -\varphi')$. After evaluation of the
$z$-derivative of $F$ we get

\[
V(\rho, \varphi, z)=\frac 1{2\pi }\int_0^s \int_0^{2\pi}
\frac{zV_0\rho' d\rho' d\varphi'}
{(\rho^2 +\rho'^2 - 2\rho\rho'\cos\gamma + z^2)^{3/2}}.
\]
The surface charge is given by the expression

\[
\sigma =-\frac{1}{4\pi }\frac{\partial V(\rho ,\varphi ,z)}{\partial z}
\Big|_{z=0} ,
\]
which gives the integral,

\[
\sigma =-\frac{V_0}{8\pi ^2}\int_0^s\int_0^{2\pi }
\frac{\rho' d\rho' d\varphi'}
{[\rho ^2+\rho ^{\prime 2}-2\rho \rho'\cos (\varphi -\varphi')]^{3/2}}.
\]
The $\rho'$-integral can be evaluated explicitly giving
\begin{eqnarray*}
\sigma & = & -\frac{V_0}{8\pi ^2} \Bigg( 
\frac{1}{\rho} \int_0^{2\pi}\frac{d\varphi'}{\sin ^2(\varphi -\varphi')}\\
& - &\int_0^{2\pi}\frac{d\varphi'}
{\sin^2(\varphi -\varphi')\sqrt{\rho^2+s^2-2\rho s\cos(\varphi-\varphi')}}\\
& +& \frac{s}{\rho} \int_0^{2\pi}
\frac{\cos (\varphi -\varphi') d\varphi'}
{\sin^2(\varphi -\varphi')
\sqrt{\rho^2+s^2-2\rho s\cos(\varphi-\varphi')}}
\Bigg).
\end{eqnarray*}
The first integral is zero while the last two integrals are elliptic
in nature and are given in tables. Here we are interested in the
long-distance behavior and thus expand the integrals in the index of
the elliptic integrals. We can then obtain the corresponding
capacitance, defined as charge divided by voltage and the volume of
the disk of thickness $h$, as
\begin{eqnarray}
C_D &=&  -\frac{1}{8\pi^2h} \Bigg(
\frac {1}{\rho^2s -s^3} + \frac{4\rho^3}{(\rho-s)^4(\rho ^2-s^2)} \nonumber\\
&+&\frac{6\rho^4s^2(4\rho^2 - 2s^2 - 2s\rho)}{(\rho - s)^9(\rho^2 -
s^2)} \nonumber\\
&+& \frac{8\rho^6s^4(6\rho^2-4s^2-2s\rho)}{(\rho-s)^{13}(\rho^2-s^2)}
+\dots\Bigg).\label{C_D}
\end{eqnarray}
One of the advantages of this result is that it shows that there are
different algebraic contributions depending on the $\rho$-range. We
show in Fig.~\ref{caps_power} that for small $\rho $ this expression
is almost linear in the log--log plot but not exactly so for larger
distances.

All the results for the full capacitance analyses are shown in
Fig.~\ref{caps_power}. There we see that the rates of decay are
different depending on the approximation but the basic nature of the
result is the algebraic or nonexponential decay of the off-diagonal
elements of the capacitance matrix. Note, however, that the
comparisons between capacitive models here are qualitative since the
definition of the parameters in the different models is not exactly
the same. We conclude that the two-disk model contains the essence of
a full capacitance matrix and thus have done most of our calculations
using this model.

\subsection{Calculational approach}

In solving the dynamical equations of motion a good amount of time is
spent in calculating the inverse of the capacitance matrix. Inversion
of $C_{\vec{r},\vec{r}'}$ may be tackled in different ways, depending
on how much we know about it.

In the case when $C_{\vec{r},\vec{r}'}$ is tri-diagonal, it has only
mutual or mutual+self capacitances and its Fourier representation in
momentum space is also tri-diagonal. Taking advantage of this form, we
can use an optimized (lower-upper) LU decomposition
algorithm\cite{fastcap} to invert $C_{\vec{r},\vec{r}'}$, such that
the calculation time grows like $\sim{N\log N}$, rather than
$\sim{N^2}$. In the case when $C_{\vec{r},\vec{r}'}$ is dense, we can
still use the translational invariance to go to momentum space,
although the multiplication (again by LU decomposition) will now grow
like $\sim{N^2}$.

Another possibility is when we already know the inverse of
$C_{\vec{r},\vec{r}'}$ and we just need an efficient way to carry out
the multiplications. Once again, translational invariance can be used
to reduce the complexity of this task. We can rewrite equation
Eq.~(\ref{eq:firstorderB}) using the definition $s(\vec r)\equiv
i_{\text{ext}}(\vec{r})
-\sum_{\vec{r}'}G^{-1}_{\vec{r},\vec{r}'}z_{\vec{r}'} 
- \sum_{\hat{\mu}}\sin \left(
\theta_{\vec{r}+\hat{\mu}}-\theta_{\vec{r}}\right)$, as
\begin{equation}
\dot{z}_{\vec{r}}=\frac {1}{\beta_c}
\sum_{\vec{r}'}C^{-1}(\vec{r},\vec{r}') s(\vec{r}').
\end{equation}
As pointed out by Phillips {\em et al.}\cite{phillips}, this has the
form of a convolution, and can thus be evaluated in $\sim N^2$
multiplications, plus the cost of performing the Fourier transform
\hbox{($\sim N\log~N$)}, as opposed to $\sim N^4$ for the direct
multiplication. Because we do not have periodicity in the current
direction, in order to obtain a linear convolution\cite{num_rec} we
use the standard `zero-padding' techniques as is done in digital
signal processing. The integration in time is carried out mostly by
the second-order Runge--Kutta (RK) algorithm.

\subsection{ Physical quantities calculated}
One of the quantities we are interested in computing is the total
voltage drop per junction in the array given by the Josephson
relation,
\begin{equation}
\langle V(t)\rangle \equiv \frac{2e}{h\nu }\frac {1}{L_x(L_y-1)}\sum_{i=1}^{L_x%
}\left( \dot{\theta}_{i,L_y-1}-\dot{\theta}_{i,1}\right).
\end{equation}
The Josephson-junction arrays studied before exhibited steps of two
very different origins: the integer steps, due to coherent
phase-slippage, and the fractional steps, due to vortex motion. It is
then important to see whether or not there are vortices associated
with possible quantized fractional steps in the I-V's. We define two
types of dual lattice vorticities: one topological and one current
related. The topological vorticity is defined as
\begin{equation}  
\label{eq:vorticitydefn}
n(\vec R) = \frac {1}{2\pi}\text{nint}\left( \sum_{P\{\vec R\}} 
\Delta\theta\right),
\end{equation}
and the current vorticity by
\begin{equation}
c(\vec R) = \sum_{P\{\vec R\}} \sin\left(\Delta\theta\right).
\end{equation}
Here $P\{\vec R\}$ denotes the sum around the four bonds enclosing the
plaquette at position $\vec R$, and $\text{nint}(x)$ is the nearest
integer to $x$. It is important to remember that the phase difference
$\Delta\theta$ is restricted to the range $[-\pi, \pi)$. The
fractional steps which we have found in the capacitive models are not
vortex related but it was important to calculate the vorticities to
check that. We do not show figures since there is nothing to see.

In particular in {\it Model C} we find that there are neither
topological vortices nor eddy current vortex produced in the
fractional steps. This appears to be true even when we let the system
evolve for 500 periods of the driving current, long after the voltage
has stabilized. In addition, in the {\sc fastcap} model, when we look
at the eddy vortices, we find that, apart from a short-lived
transient, no vortices are present.

To further understand what is happening we also have looked at the
spectral function defined as
\begin{equation}
S(\nu ) = \lim_{\tau\rightarrow\infty}{\Bigg| {1\over \tau}{\int_0^\tau
V(t)
e^{i2\pi \nu t}dt}\Bigg|^2}.
\end{equation}
For integer steps $S(\nu)$ has peaks at frequencies $n\nu$, for
periodic and disordered arrays\cite{numshap1}. This is consistent with
the resonant nature of the integer steps. For the disordered arrays
that produce the ACVS state there are half-integer vortex steps,
corresponding to an $S(\nu )$ that has additional peaks at frequencies
$n\nu /2$.

\section{Results}

\label{sec:numerics}
In this Section we consider the response of {\em Models A--C} under
different current-driven conditions. First we consider the case when
the array is subjected to a DC external current and next when there is
a DC+AC current drive. We also look at other physical measures defined
above, that further help us in the analysis of the results.

\subsection{DC current--voltage characteristics}
Josephson-junction arrays driven only by a DC current have been
studied before in the overdamped case and also in {\it Model
A}\cite{rev}. Here we examine the I-V's for $8\times 8$ arrays in {\it
Model C} based on the two-disk approximation as well as a square plate
model based on a {\sc fastcap} calculation.

We have computed the DC I-V's for {\it Model C} in a DC drive for
different values of the parameter, $D$. When $D\lesssim 1$, and with
$\Lambda _0\gg L_x,L_y$, there is basically no screening and thus the
I-V's have the usual quadratic form\cite{modb}:
\begin{equation}
V=\sqrt{i_{dc}^2-1}.
\end{equation}
In Fig.~\ref{dc_iv} we show the behavior of the I-V's for two extreme
values of the disk diameter $D$. As $D$ slowly decreases from 1.0, we
find little change in the shape of the I-V's until $D\sim 0.7$. At
this point, the I-V rather loses its smooth curved shape, and assumes
a piecewise linear form characterized by a very sudden increase, and
then to linear ohmic behavior as the current is increased. We also
note a slightly lower critical current for the latter case.

Next we also calculated the I-V in the {\sc fastcap} disk model
considered in {\bf Ref}.~\onlinecite{whan} in a r\'egime close to that of
{\it Model C} considered above. In this case the results shown in
Fig.~\ref{FASTCAP_dc} are somewhat more tentative, for we may be
operating outside the r{\'e}gime in which the {\sc fastcap} model is
valid. Qualitatively, the I-V's display the transition from a smooth
shape for a large value of $\tilde D$ (defined as a normalized diameter of
the disk) to piecewise linear at
smaller $\tilde D$-values. Note that quantitatively this length 
$\tilde D$ is not exactly the same as the one defined for {\it Model C}. The
conclusion from these calculations is, however, that the results
evaluated with both models are qualitatively, if not quantitatively
similar.

\subsection{DC+AC current--voltage characteristics}
\label{sec:numerics_ac} 

Here we present results for DC+AC driven capacitive JJA and compare
results for {\em Models A}, {\em B}, and {\em C}. {\it Model A} has
been studied to some extent and we briefly mention those results here,
but we will be mostly concerned with results for {\it Models B} and
{\it C}.

\subsubsection{Model A: Mutual capacitances only.}

\label{subsec:modelA} 
This case has been studied in {\bf Ref}.~\onlinecite{hagenaars} mostly in
the transition region between regular and chaotic behavior for the
single-junction problem. Apart from the integer giant Shapiro steps
Hagenaars {\em et al.}\cite{hagenaars} find the formation of a
half-integer step, within a range of values of $\beta_c$ which include
chaos in the single junction, with characteristics similar to those of
the ACVS state mentioned in the introduction. In our analysis we
mostly stay away from the single-junction chaotic state, and do not
see the half-integer steps. Next we look at what happens when there is
screening in the capacitive models.

\subsubsection{Model B: Self+mutual capacitances only}

\label{subsec:modelB} 
Here we present results when the capacitance matrix is composed of the
self capacitance of each island to ground plus the mutual capacitance
of each island to {\em only} its nearest neighbors. The time-step used
in the second-order Runge--Kutta calculations was $0.05$ for all the
results in this section, with a current grid of $\delta i_{dc}/{i_c} =
0.005$. Except where otherwise noted, averaging was carried out over
$6\times 10^4$ time-steps, after throwing out the first $2\times 10^4$
for equilibration.

As mentioned before, addition of the self capacitance to {\it Model A}
introduces exponential screening that for large distances is measured
by the screening length $\Lambda_0=\sqrt{C_m/C_s}=1/\sqrt{r_c}$. Our
results were obtained principally on an $8\times8$ array, but we have
seen similar behavior in much larger arrays (e.g.,\ $40\times40$, see
Fig.~\ref{lx40_steps}). We find that there are new giant
capacitive fractional steps (GCFS) as we vary $\Lambda_0$. When such
steps occur, they appear as multiples of $\frac{n}{L_y-1}$, where $n =
1, 2, 3$ and possibly 4.

The results obtained when $\Lambda_0\gg \,L_y$ were, as must be,
identical to those obtained with the capacitance matrix composed of
{\it Model A}. The existence of these integer-steps depends on the
locked-in coherent phase oscillation of all junctions in the array
with the drive. On reducing the screening length to a value smaller
than the array size, the integer-steps began to disappear, and then
new steps appeared at fractional values of the normalized voltage.

It can be seen in Fig.~\ref{stepdisappear} that, for $\Lambda_0 >
L_y$, the width of the integer-step is entirely unaffected by
screening. As $\Lambda_0$ approaches $L_y/2$, the step at $n=1$ begins
to shrink, and as $\Lambda_0 \rightarrow 1$, the step is rapidly
destroyed. In Fig.~\ref{screen} we show the change in the width of the
$n=1$ step as a function of $\Lambda_0$.

We then examined the step width for the GCFS $\frac{2}{7}$-step as
functions of the frequency $\nu$, $\beta_c$ and $\Lambda_0$. As can be
seen from Fig.~\ref{width_nu}, the appearance of this step is quite
strongly dependent on the drive frequency, and it is absent for
frequencies 15\% greater than or less than $\nu =0.1$, with $\beta_c =
0.50$. Next we fixed $\nu$, and looked at the width of the fractional
steps, as a function of $\beta_c$. Our results are shown in
Fig.~\ref{width_betac}. The $\frac{1}{7}$- and $\frac{2}{7}$-steps are
seen to appear for $\beta_c \sim 0.3$, reach their maximum values at
around $\beta_c\sim 0.7$, and then slowly disappear, as $\beta_c$ is
increased. All fractional and integer steps have disappeared by the
time $\beta_c$ has reached the value 3.0. Each step has a {\em maximal
screening length} denoted $\Lambda_0^{\text{max}}$ for which its width
is maximized, and the lower the order of the step, the smaller its
$\Lambda_0^{\text{max}}$. We mostly stayed outside the range of
$\beta_c$ for which the single-junction behavior is chaotic, though
Fig.~\ref{width_betac} shows that chaos is not necessary to generate
the GCFS.

We then fixed both $\nu=0.1$ and $\beta_c=0.6$ (parameters for which
the step-widths appear to be maximized), and looked at how the
step-width varies as a function of $\Lambda_0$. The results are shown
in Fig.~\ref{width_lambda}. We decreased the screening length from a
large value until $\Lambda_0$ had reached half the array size, i.e.,
$\Lambda_0 \sim 4$. At this point, the fractional step size jumps up
rather rapidly to almost its maximum value, remains there until the
screening length has reached the value 1.0, and then falls rapidly to
zero once again. This behavior can be seen to happen for both the
$\frac{1}{7}$- and $\frac{2}{7}$-steps.

We obtained these steps by starting up from a lower value of $i_{dc}$,
since one cannot see the step by simply setting the value of $i_{dc}$
according to the expected step values in the I-V's. The fractional
steps are hysteretic, as indicated by the inset in
Fig.~\ref{width_lambda}. This is a typical property of {\it Model A}
arrays, and has been seen in previous studies, whose authors interpret
them as evidence of vortex inertia\cite{modb}. Both of these facts
indicate that the GCFS are metastable, since their existence depends
on the previous history conditions of the array.

Usually, the appearance of a step at a voltage $V = \frac{p}{q}$
indicates that a process with frequency $\frac{p}{q}\nu$ underlies the
production of the step. This should be visible in the Fourier power
spectrum of the array voltage. For the integer steps that is what we
see but not so for the GCFS, which signals that they are of a
different nature to the steps studied before, as we further discuss
below.

\subsubsection{Model C: Full capacitance matrix}

\label{subsec:modelC} 
In this subsection, we present our results for the capacitance matrix
composed of all nonzero elements. In this case, the RK time-step
calculation used was $0.05$ for all the results in this section, with
a current grid of $\delta i_{dc}/i_c = 0.005$. Except where otherwise
noted, averaging was carried out over $6\times 10^4$ time-steps and
the first $2\times 10^4$ were used for equilibration. Longer
time-series have shown that while there are some long-lived
transients, eventually the state becomes stable.

Since the off-diagonal matrix elements of $C_{\vec{r},\vec{r}'}$ decay
algebraically, the definition of screening length as done in {\it
Model B} does not strictly apply. Nonetheless, because of the specific
decay in {\it Model C} we can still quantify the screening by the
ratio of the off--diagonal capacitances to the diagonal ones, using
the definition of {\it Model B}. We could have used another definition
in which we sum over all the matrix elements, normalized
appropriately, but the end result, that is the measure of screening,
would be about the same. (In fact, we did just that, and found that
while it resulted in a rescaling of the screening length by a small
numerical factor, it did not change the qualitative behavior.)

Our results were obtained principally on an $8\times8$ array, but we
have seen similar behavior in much larger arrays (e.g.,
$40\times40$). In this model, in contrast to {\it Model B}, we have
seen only the $\frac{2}{7}$-step which has to do with the nonzero
decay of the off-diagonal matrix elements of the capacitance matrix.
In {\it Model C} we can move continuously from {\it Model A} to {\it
Model B} by adjusting two parameters: the self capacitance (by varying
$r_c$) and the diameter of the disk, $D$, expressed in units of the
lattice spacing. $D$ takes values in the range $[0, 1)$, the upper
limit giving {\em Model B} (or {\it Model A} if, in addition,
$r_c=0$). The lower limit corresponds to an array of superconducting
dots which are too far separated for tunneling. As we decrease $D$
from $1$, for $\Lambda_0\gg L_x$, a voltage step appears at $\langle V
\rangle = \frac{2}{7}$, for $D = 0.997$, and disappears for $D =
0.95$. The results are shown in Fig.~\ref{ivsD}, for $\Lambda_0 = 32$
on an $8\times 8$ array. The maximum width of the $\frac{2}{7}$-step
is of the same order of the half-step seen before\cite{hagenaars},
though we chose $\beta_c = 0.5$ specifically to be outside the chaotic
regime for which Hagenaars {\em et al.}\cite{hagenaars} saw that step.

The magnitude of the step appears to be $\frac{2}{L_y-1}$, based on
runs for $L_y = 4$, 8, and 16: that is, for each lattice size, we
observe the appearance and subsequent disappearance of a step at
$\langle V \rangle = \frac{2}{L_y-1}$ as $D$ is varied down from $1$.
The occurrence of $L_y$ in the expressions suggests that the GCFS are
edge-like effects. They do not occur for {\it Model A}, but do occur
for certain parameter values in {\it Model B}, as can be seen in
Fig.~\ref{stepdisappear}. We also found qualitatively similar behavior
(i.e., the existence of a step only for a limited range of $D \sim 1$)
using the expression for the inverse capacitance matrix model of Whan
{\em et al.}\cite{whan}

We have examined the width of the $\frac{2}{7}$-step as a function of
$\Lambda_0$, as we fixed both $\nu=0.1$ and $\beta_c=0.5$. The results
are shown in Fig.~\ref{modelC_width_lambda}. There are some clear
differences between the behaviors of {\it Models B} and {\it C}.  We
stress that because the screening in {\em Model B} is exponential for
long and logarithmic for short distances, as compared to the algebraic
decay in {\it Model C}, this is clearly seen in the different results
in this figure.

The GCFS steps appear to be related to a partially screened
phase-locked state. They do not appear to be metastable, in the sense
that there is no need to `step' up onto them, as was found in {\it
Model B}. We also produced animations that show no stable vortex
patterns, but indicate that all junctions in a transverse row evolve
in phase without an apparent simple relationship between adjacent
rows. We examined the voltage states of all junctions along the
central column of the array (that is, the ladder consisting of all the
junctions with $x=4$), and found that there are essentially two kinds
of behavior: The behavior of the edge junctions seems to be
symmetrical about the line $y=L_y/2$, with their voltage (normalized
for frequency) equal to unity. The other junctions, while not locked
in or very much out of phase with each other, are in zero-voltage
states. This is clearly due to screening.

We also looked at the spectral function of a {\it Model C} array. We
find, sitting on the $\frac{2}{7}$-step (having stepped up onto it),
that when we study the spectra of the junctions within the center row,
at one of the edges (all within the ladder defined by $x=4$) or the
whole array, all three cases show similar behavior.

There is a small but very long-lived transient, which shows up in the
power spectral density, as two peaks at frequencies approximately
equal to $\sim \frac{4\nu}{7}$ and $\frac{5\nu}{7}$ (as shown in the
inset of Fig.~\ref{modelC_edge_on}). Also present in the spectrum are
the expected drive frequency $\nu = 0.1$, and its harmonics. We have
verified that it {\em is} a transient and is not responsible for the
appearance of the fractional steps, by allowing the array to
equilibrate for very long periods of time. Under these conditions, we
still see the step, and the power spectrum is very clean and shows no
trace of the transient, as shown in the inset of
Fig.~\ref{modelC_edge_on}.

Figure~\ref{modelC_edge_on} shows the I-V's of an edge row of
junctions, an interior row and the average of the whole array. Both
edge rows behave similarly to each other, but differenatly from the
interior rows. It is clear that the edge row enters a nonzero voltage
state at a lower current than the interior row ($i_{dc} =0.81$ {\em
versus} $i_{dc} =0.885$). That we have two out of seven rows in a
voltage-producing state, and five out of seven in a zero-voltage
state, seems to indicate that the $\frac{2}{7}$-step has its origin
not in a coherent oscillation of the array as a whole, but rather in
the fact that some of the junctions get driven into a nonzero voltage
state at a lower current than others, producing the observed
$\frac{2}{7}$-step. This is more easily seen in the $16\times 16$
array, where multiple steps are visible. In particular, for $D=0.999$,
we see steps at $\frac{2}{15}$, $\frac{4}{15}$ and
$\frac{6}{15}$. Looking at the I-V's for the individual rows, we can
see first the outermost (edge) two rows (one at each boundary)
switching on, followed by the next outermost rows, followed by the
next outermost ones. The points where these rows sequentially switch
on correspond to the beginning of a new step.

\subsubsection{Size dependence}
In the results reported above the size of the system appears
prominently. The question is then how stable are these results as a
function of lattice size and parameter values. Most of the results
described in the previous sections have been generated in lattices of
size $L_x =L_y=8$. The steps we find are multiples of $\frac{1}{7}$
(since $L_y-1 \equiv 7$). We also have looked at array sizes of $L_x =
L_y = 16, 40$. In both cases, we found analogous GCFS at integer
multiples of $\frac{1}{15}$ and $\frac{1}{39}$. In {\it Model B}, we
notice that the maximum step width for the $\frac{1}{7}$-step in the
$8\times 8$ array is about $\Delta i_{dc} = 0.10$, whereas for the
$40\times40$ array, we find it to be $\sim 0.085$. The widths are
therefore comparable, and seem largely independent of lattice
size. Figure \ref{lx40_steps} shows the I-V for a $40\times40$ array,
with a screening length much smaller than the array ($\Lambda_0 = 1.75
$).

In {\it Model B} we have examined the critical screening length
$\Lambda_0^{\text{crit}}$ for which the first step appears as a
function of size $L_x$. We show in Fig.~\ref{crit_screen_len} the
corresponding data, along with an approximate fit that gives
$\Lambda_0^{\text{crit}}\sim L_x^{0.914\pm0.058}$. Figure~\ref{screen_len_max}
shows the maximum width of the first step as a function of $L_x$ for a
given maximum screening length $\Lambda_0^{\text{max}}$. Note that the
first step is generally the biggest, regardless of lattice size. In
addition, we show in Fig.~\ref{maxwidth_lx} that the maximal width of
a step of a given order appears to saturate once the lattice size is
above a certain minimum.

\begin{figure} 
\caption{Step width $\Delta i_{\text{dc}}$ as a function of disk
diameter, $D$, in {\it Model C} for a $16\times16$ array. Data are for
$\frac{2}{15}$-step (the lowest-order step seen). Parameter values are
$\beta_c = 0.50$, $\nu = 0.10$.\label{i_D_lx16}}
\end{figure}
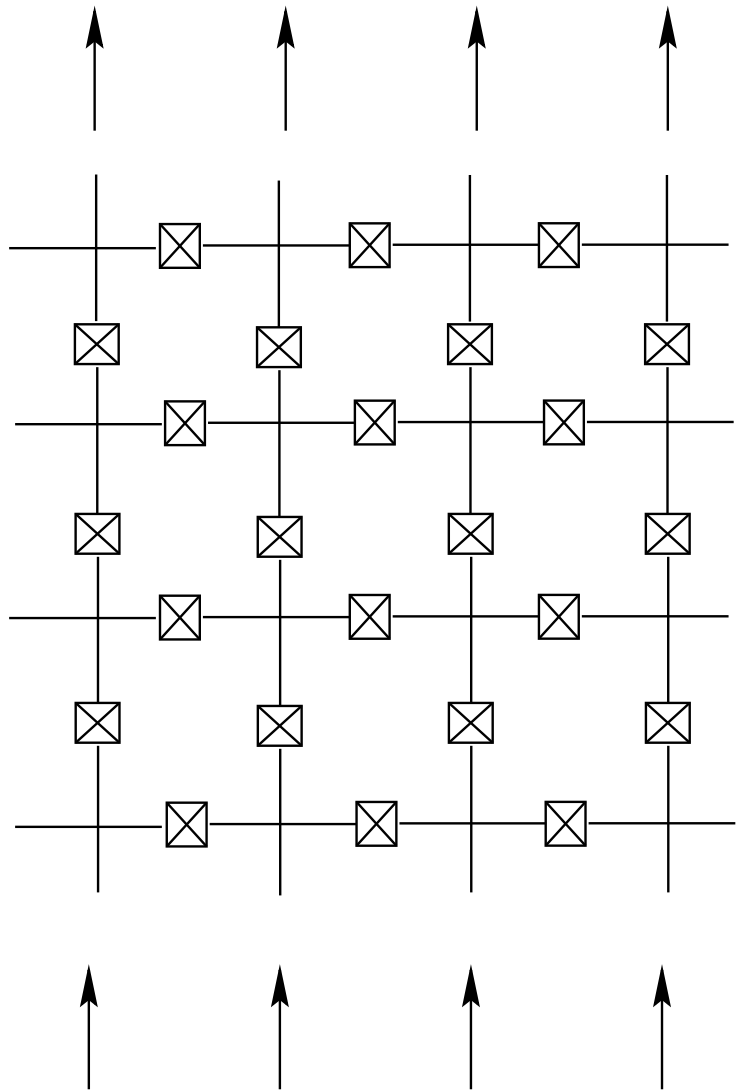
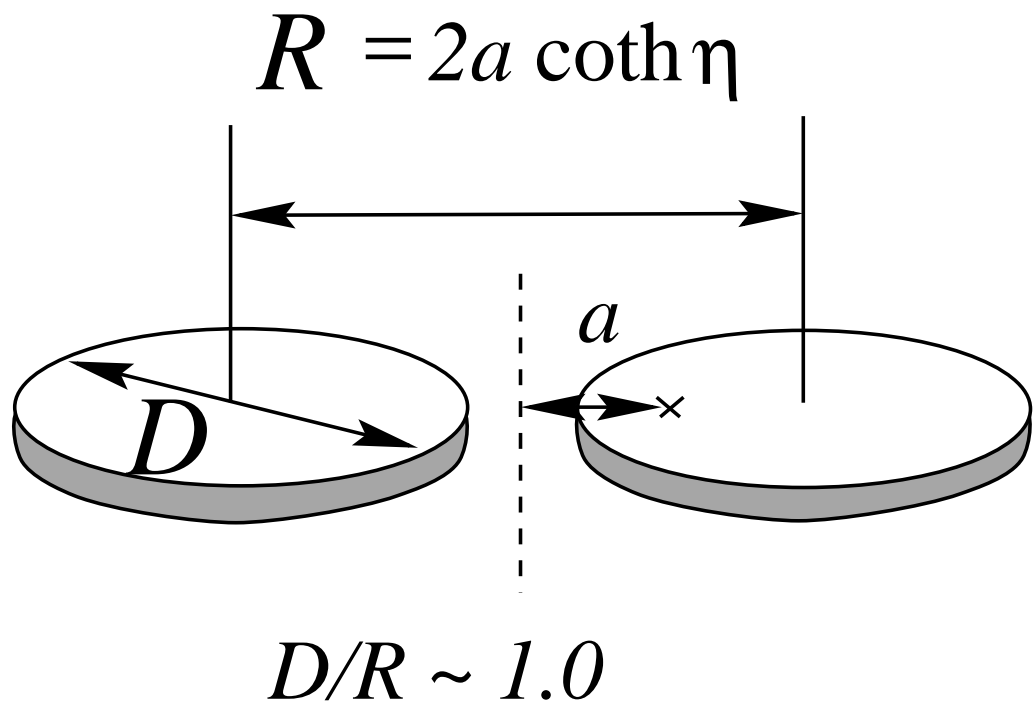
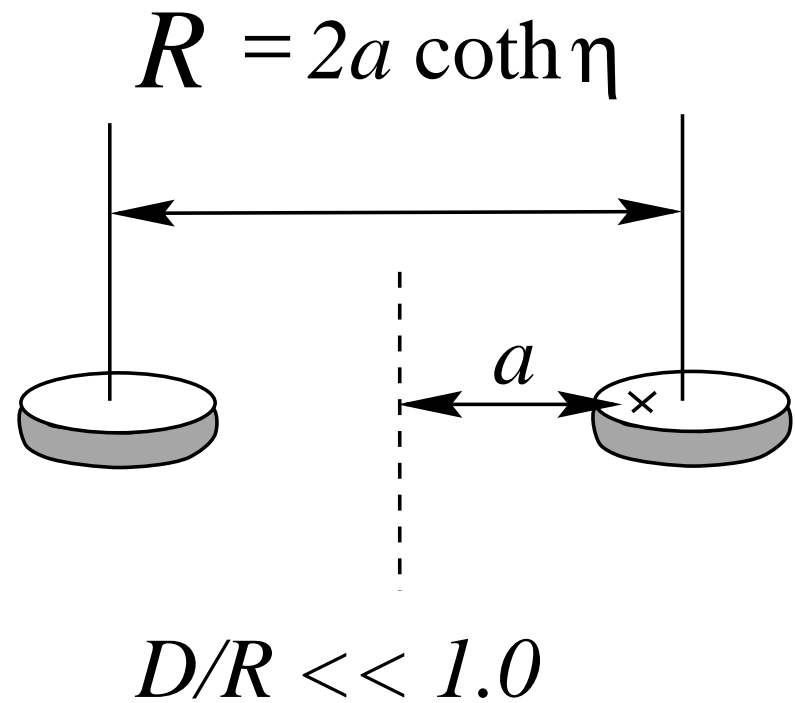
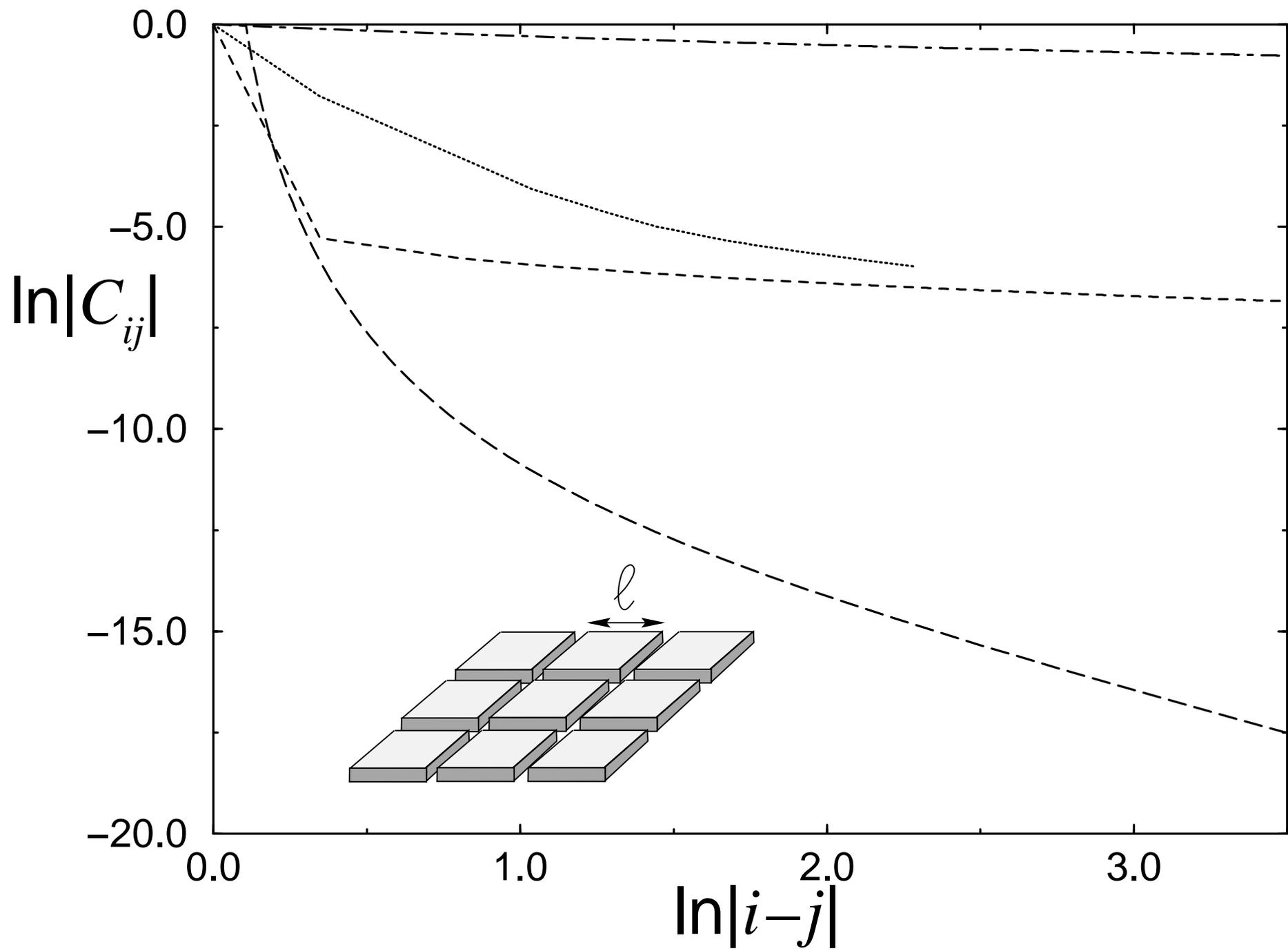
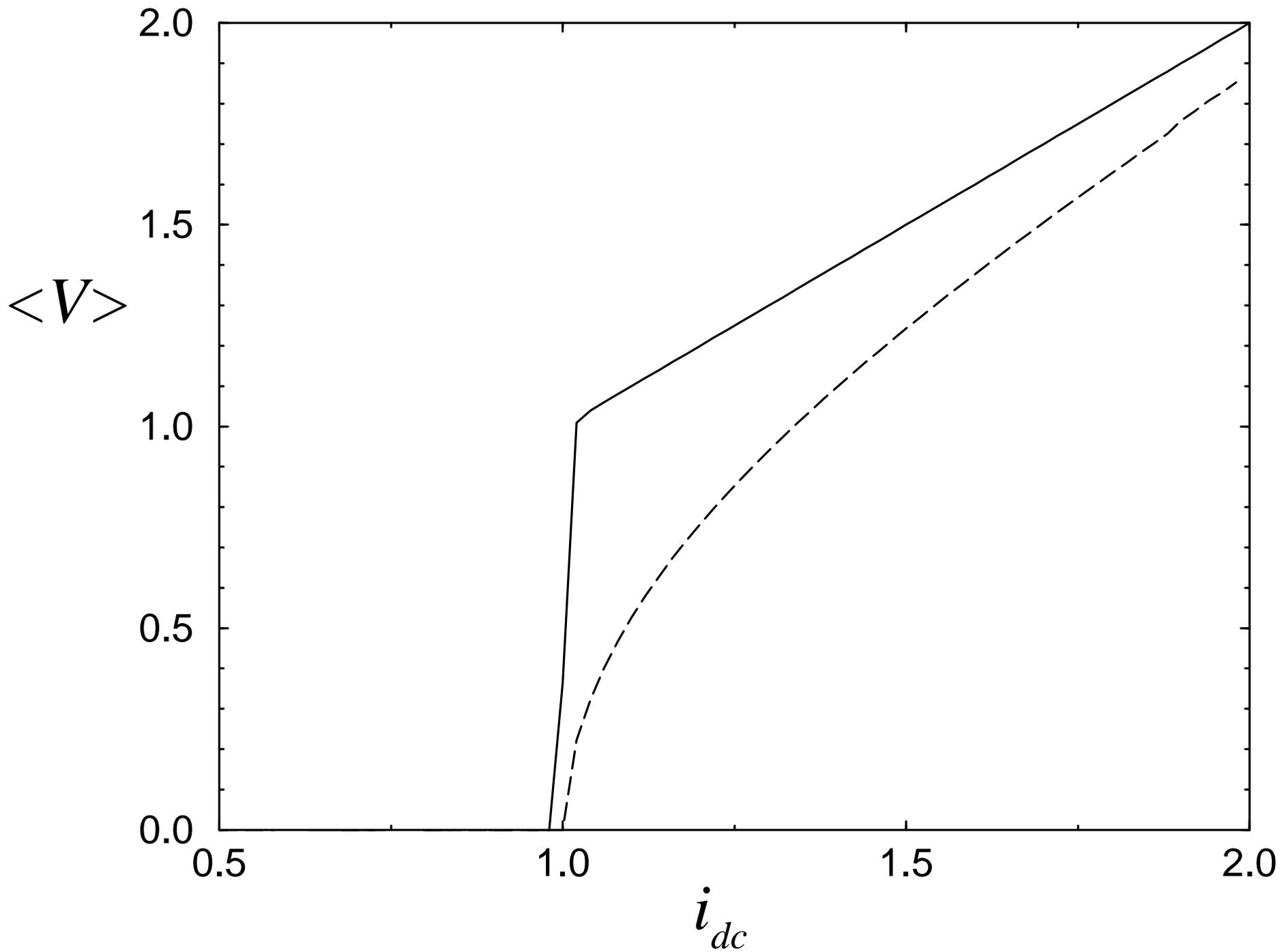
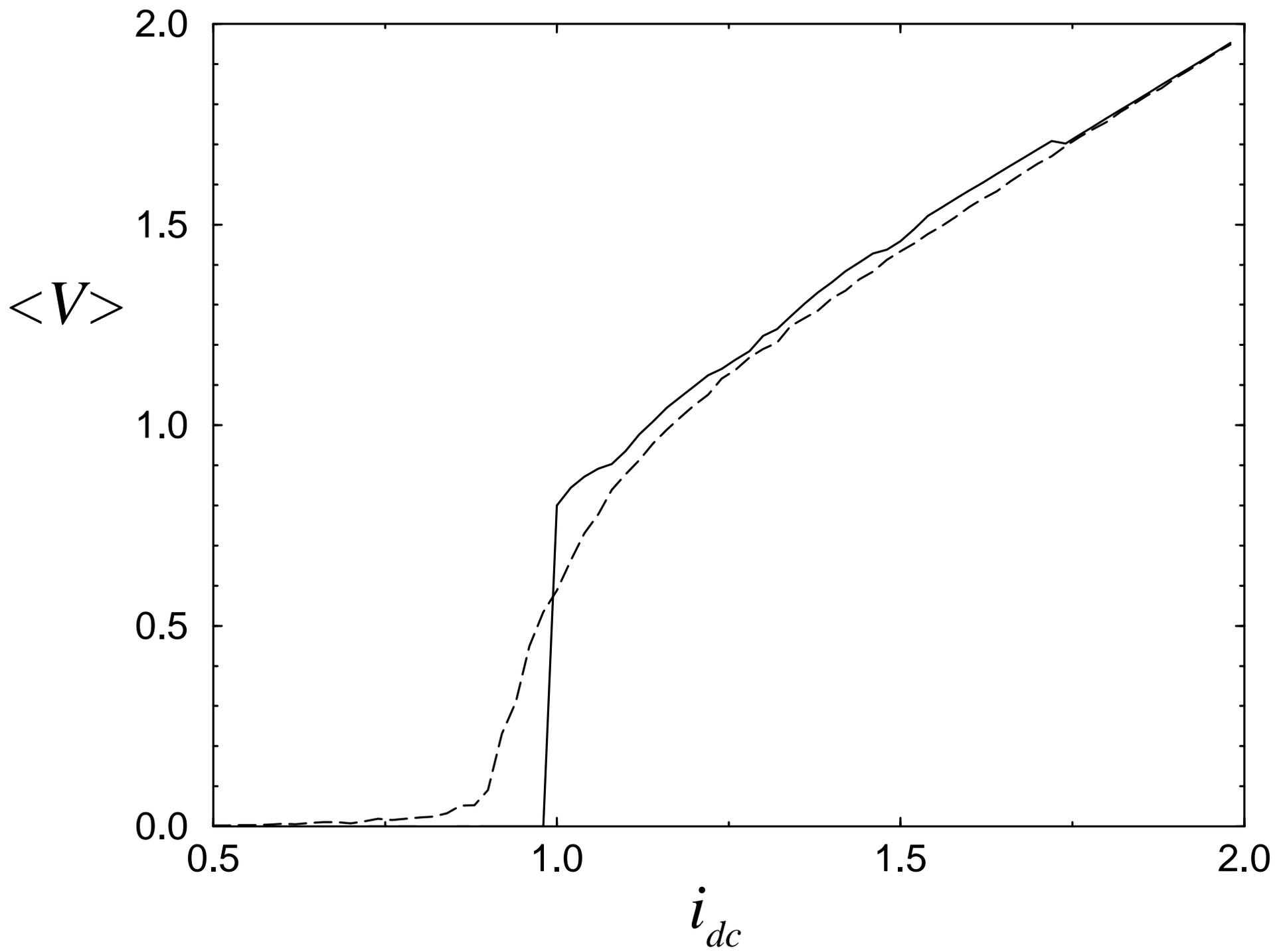
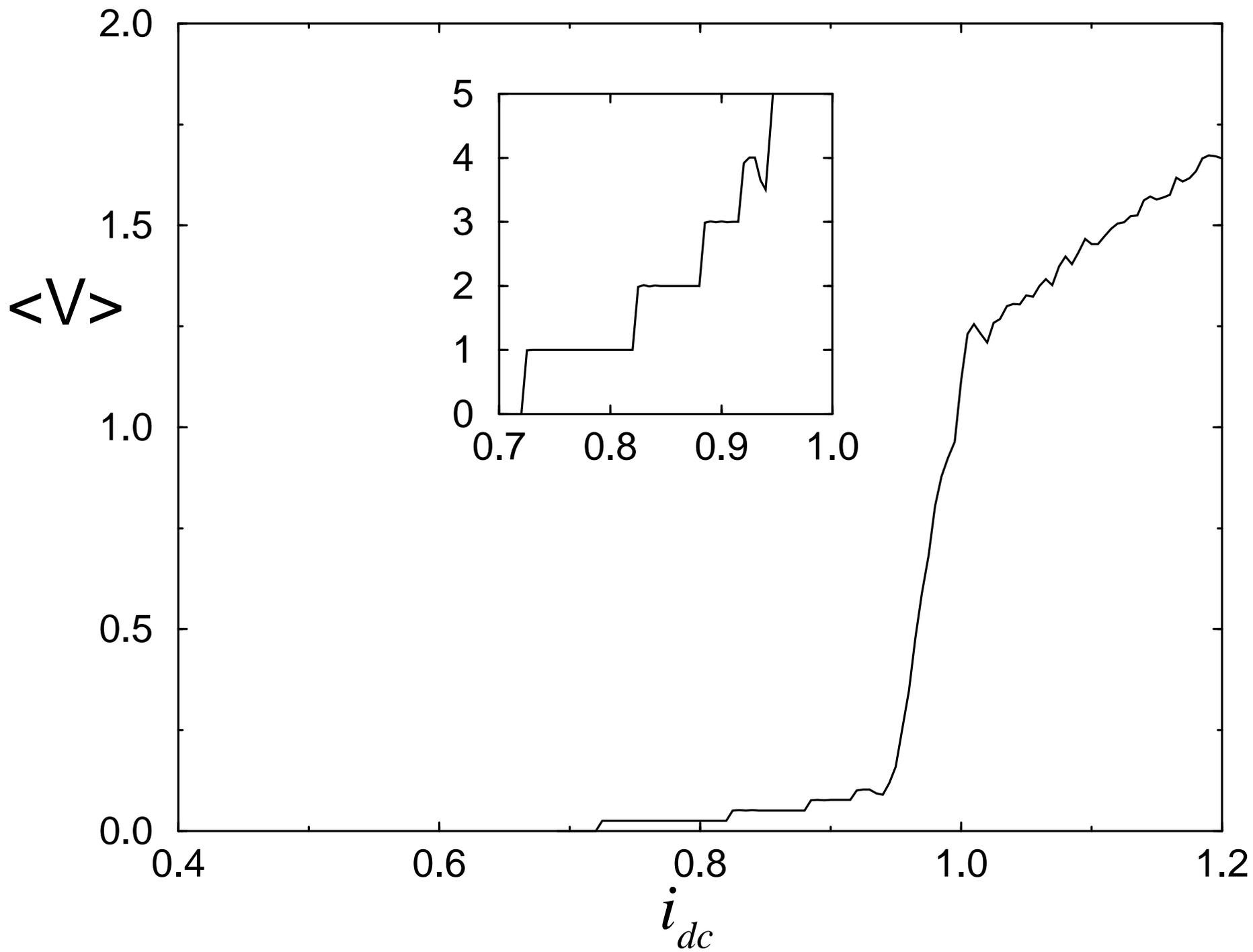
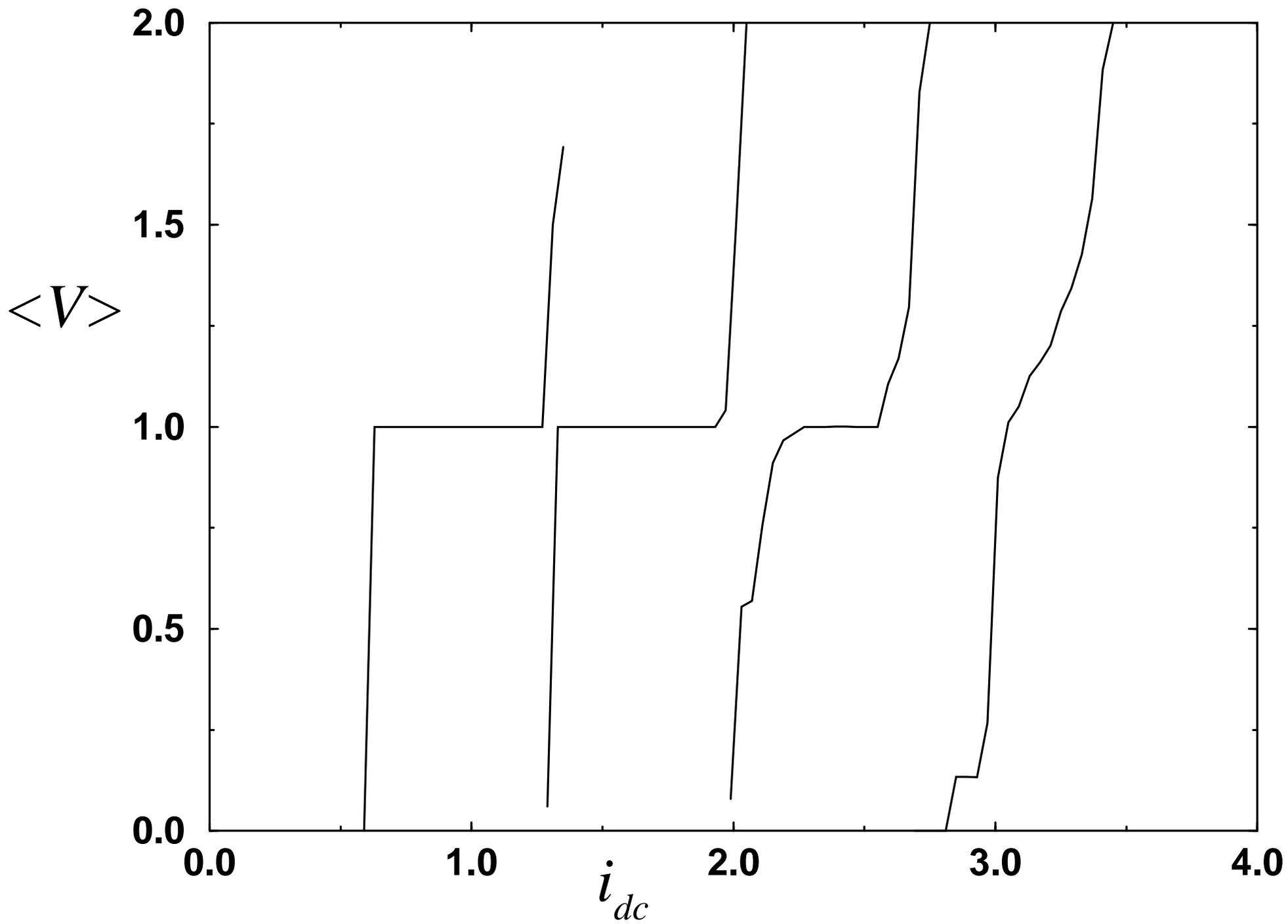
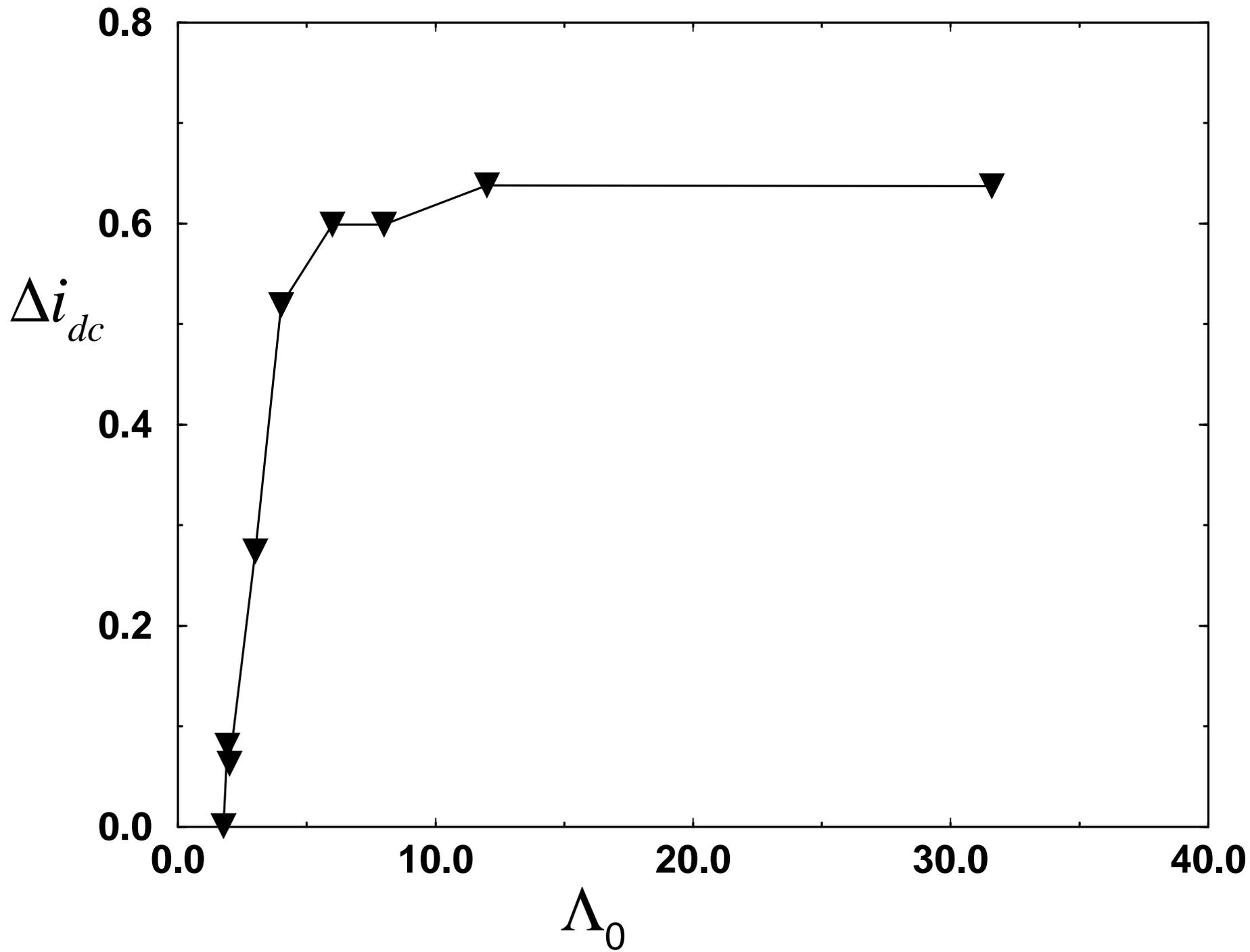
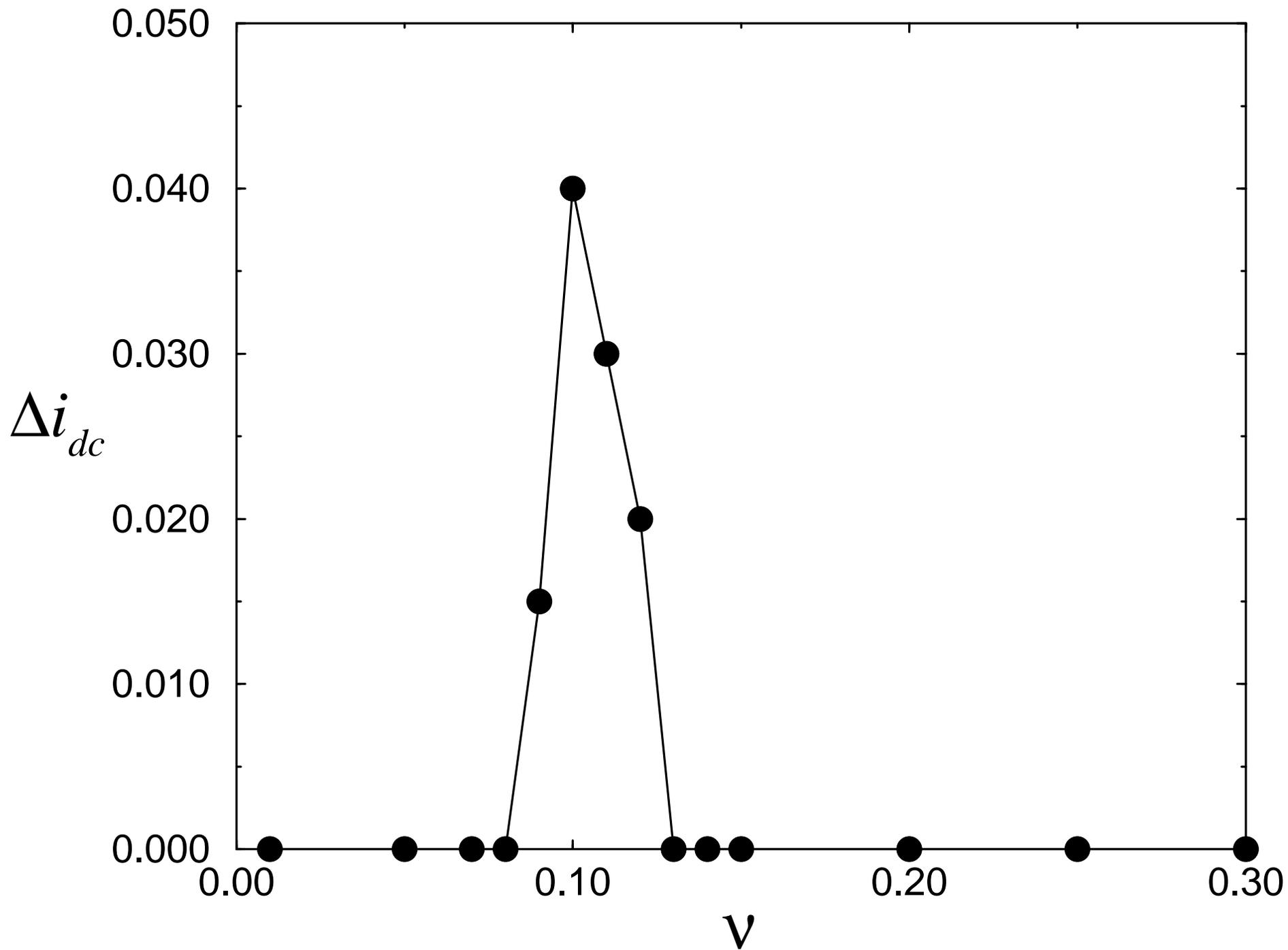
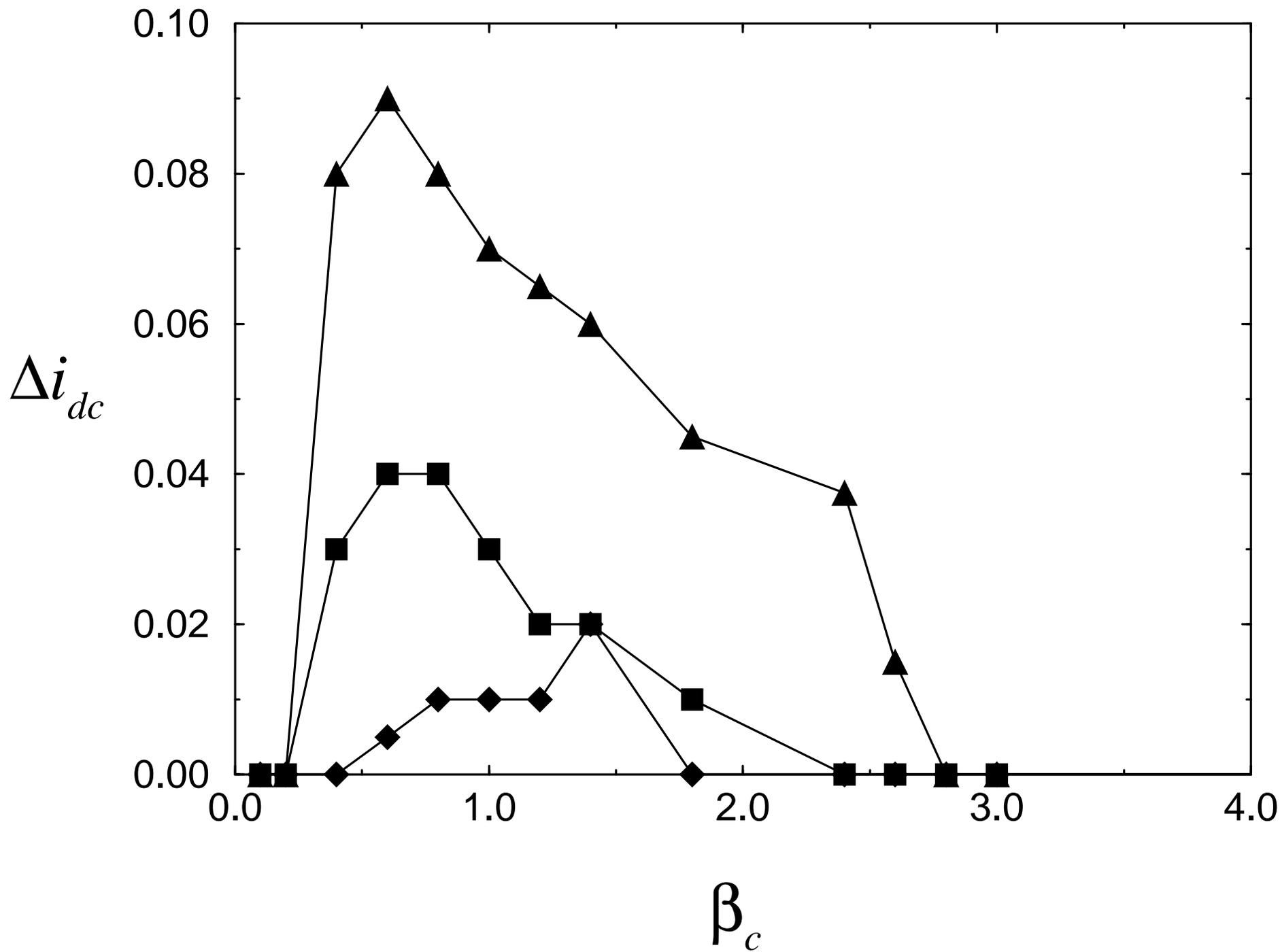
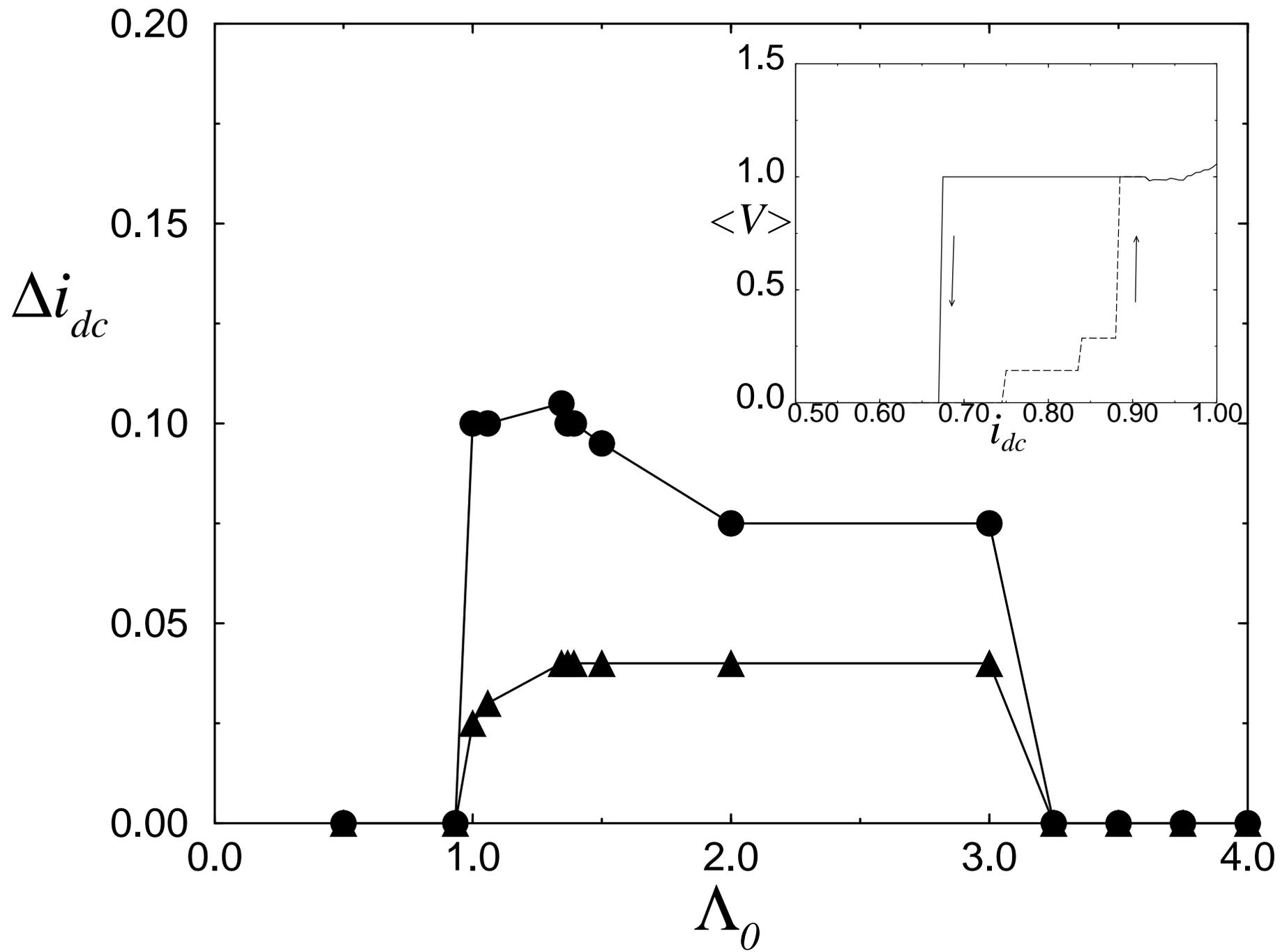
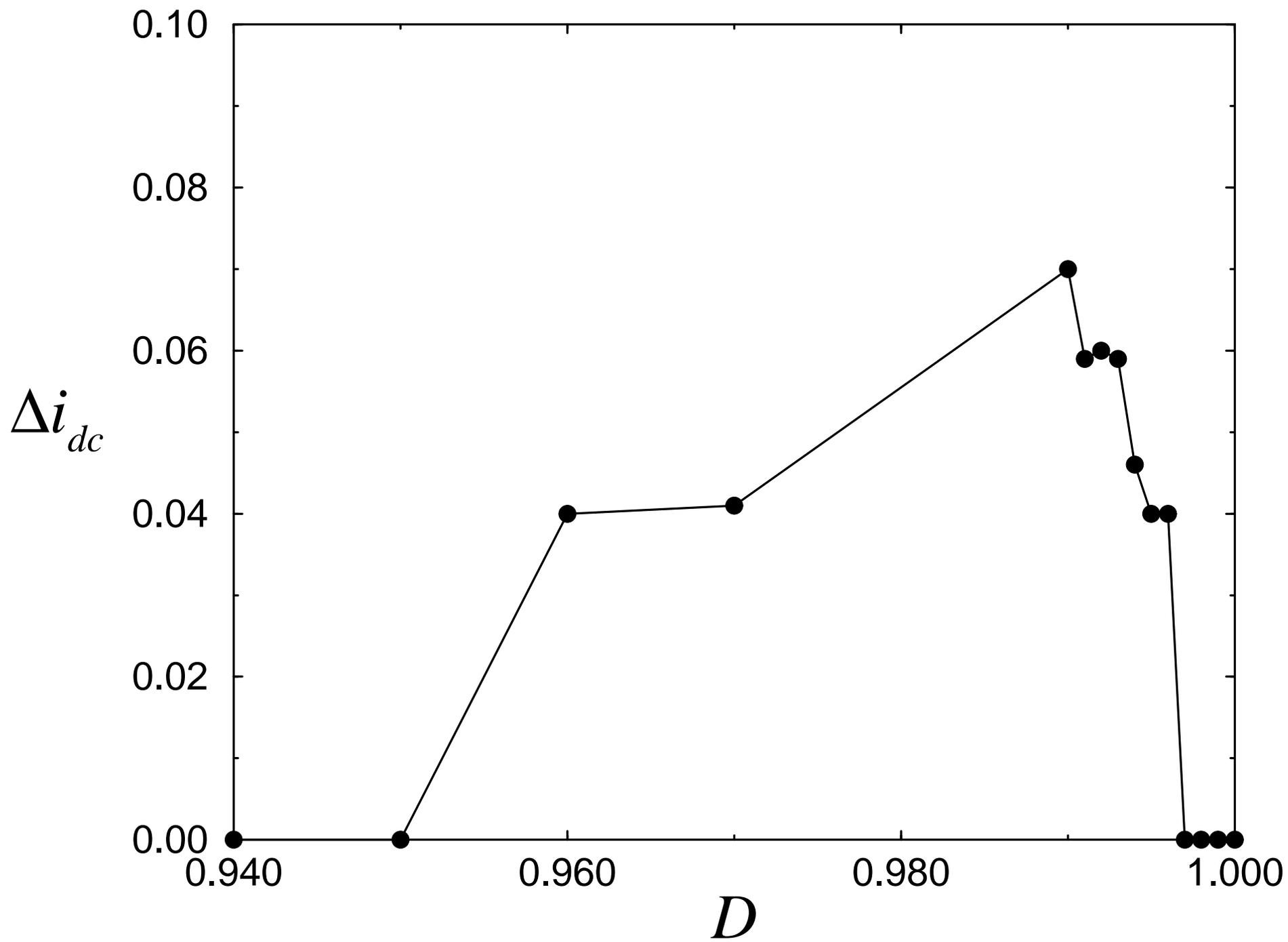
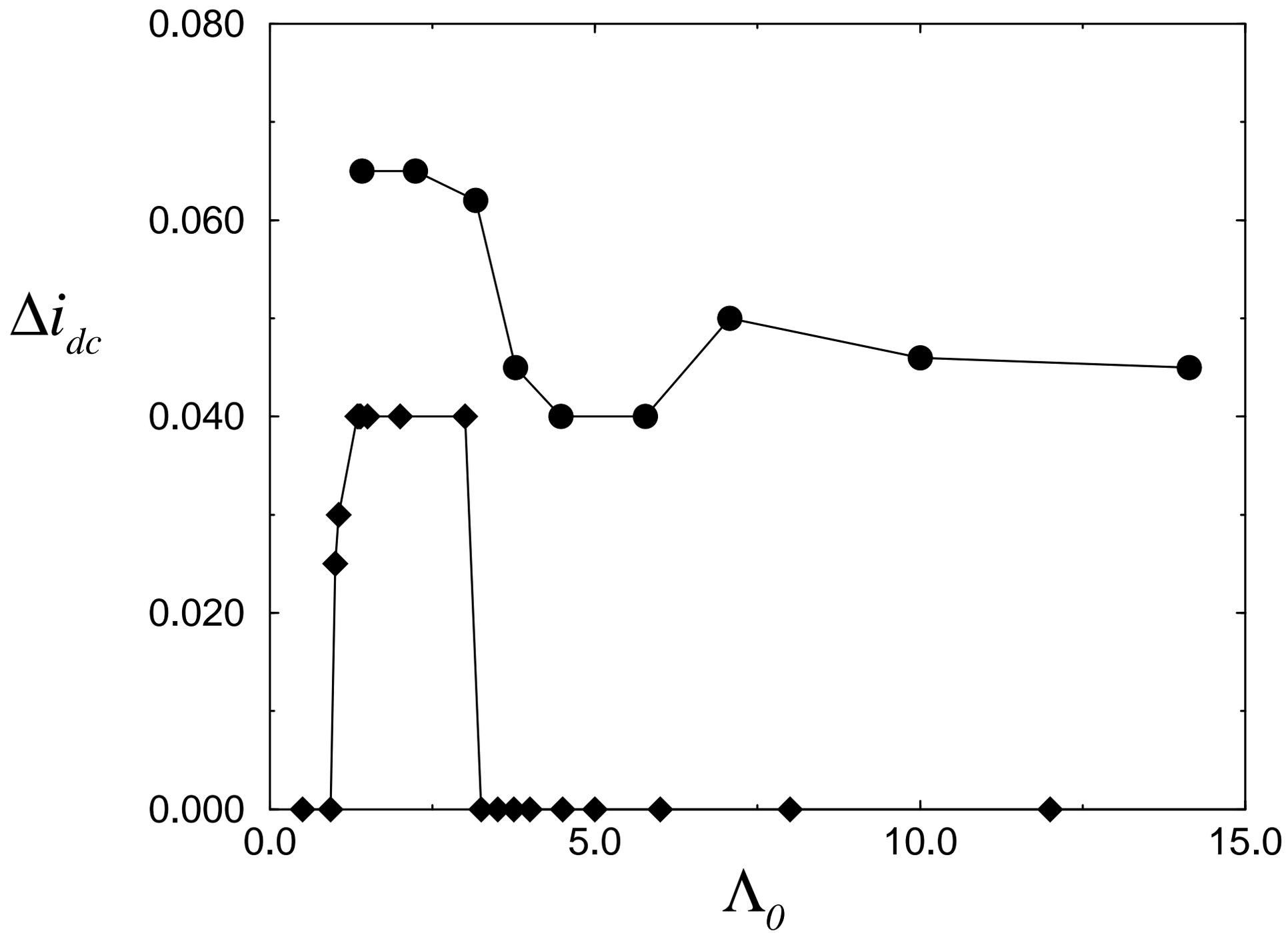
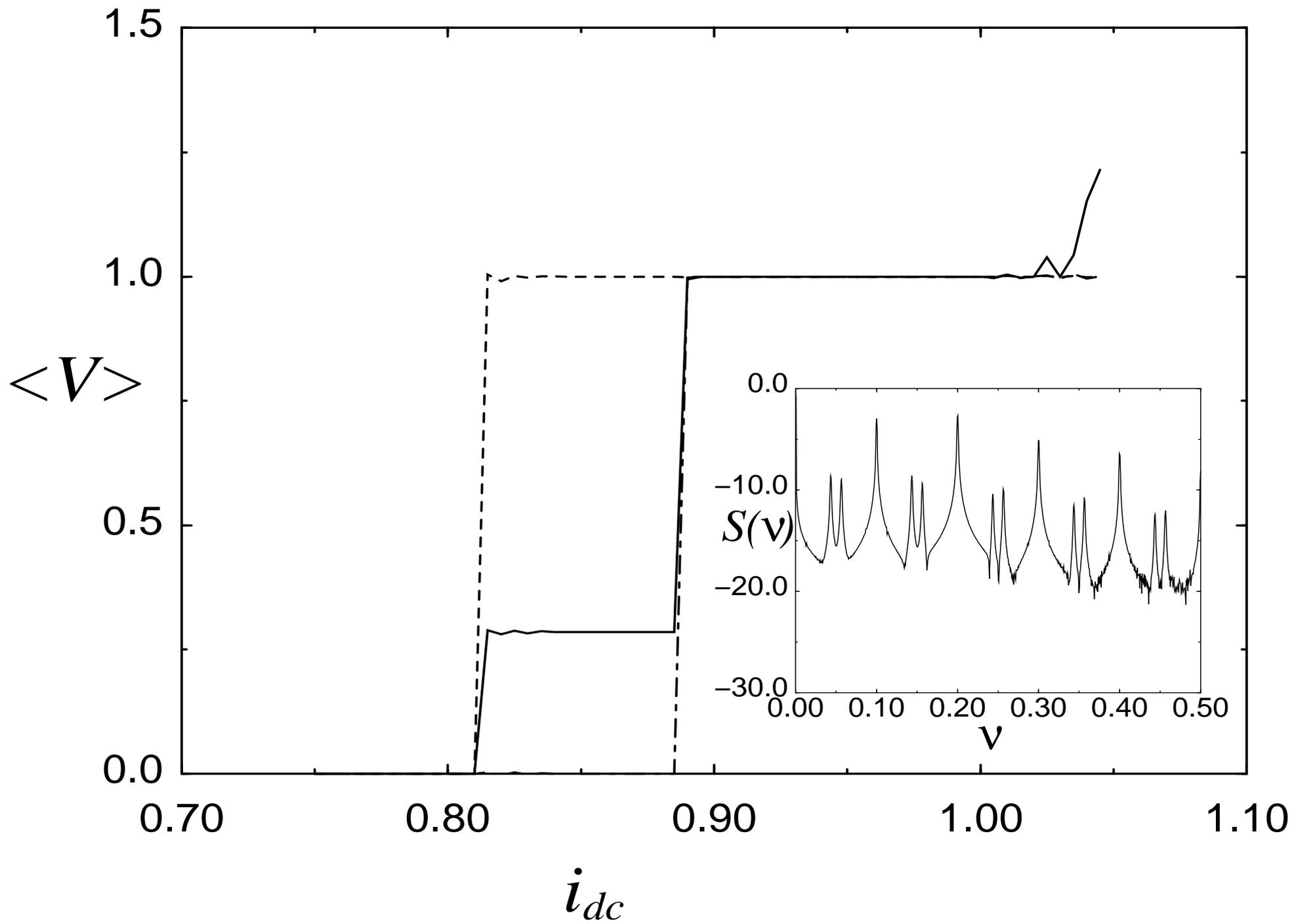
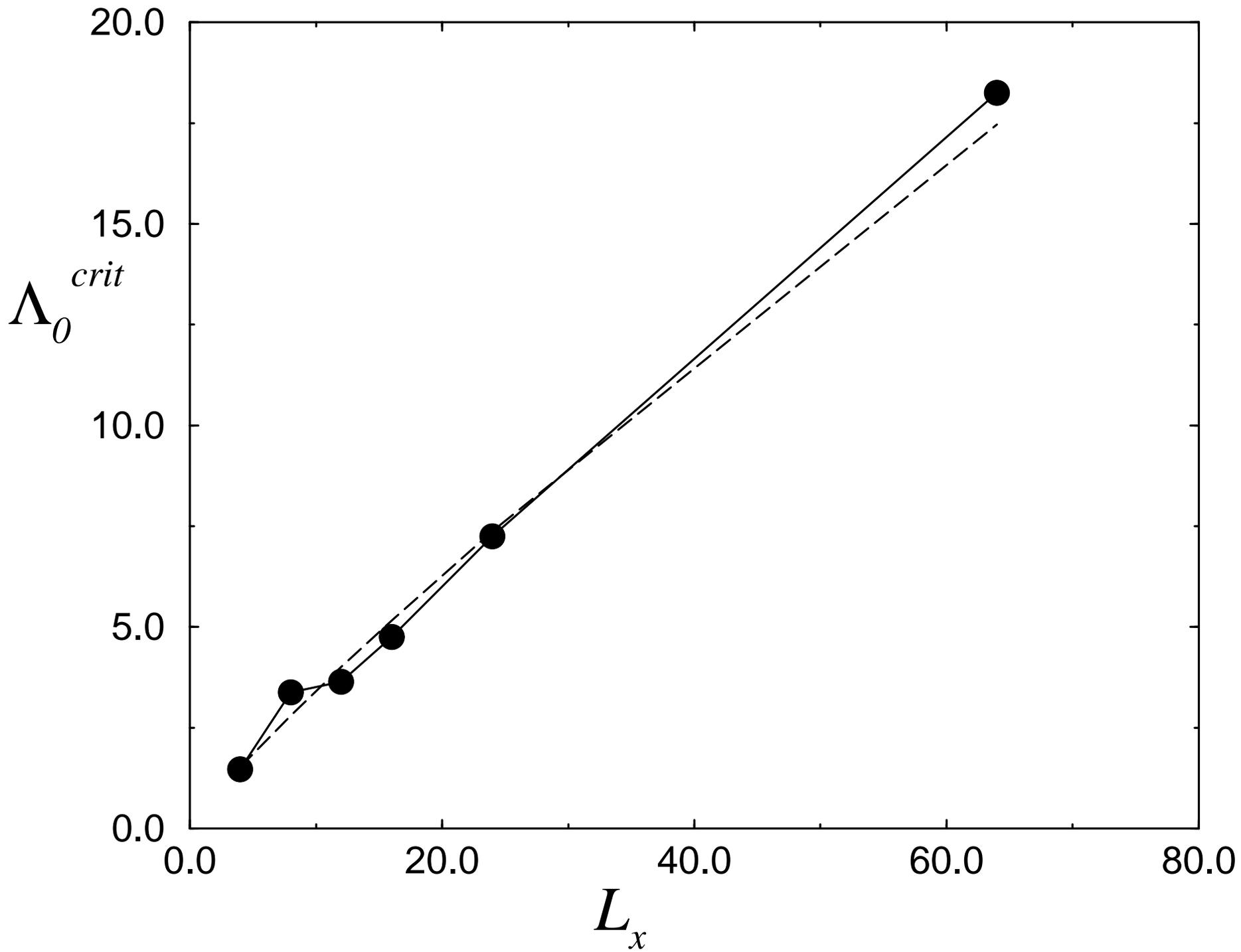
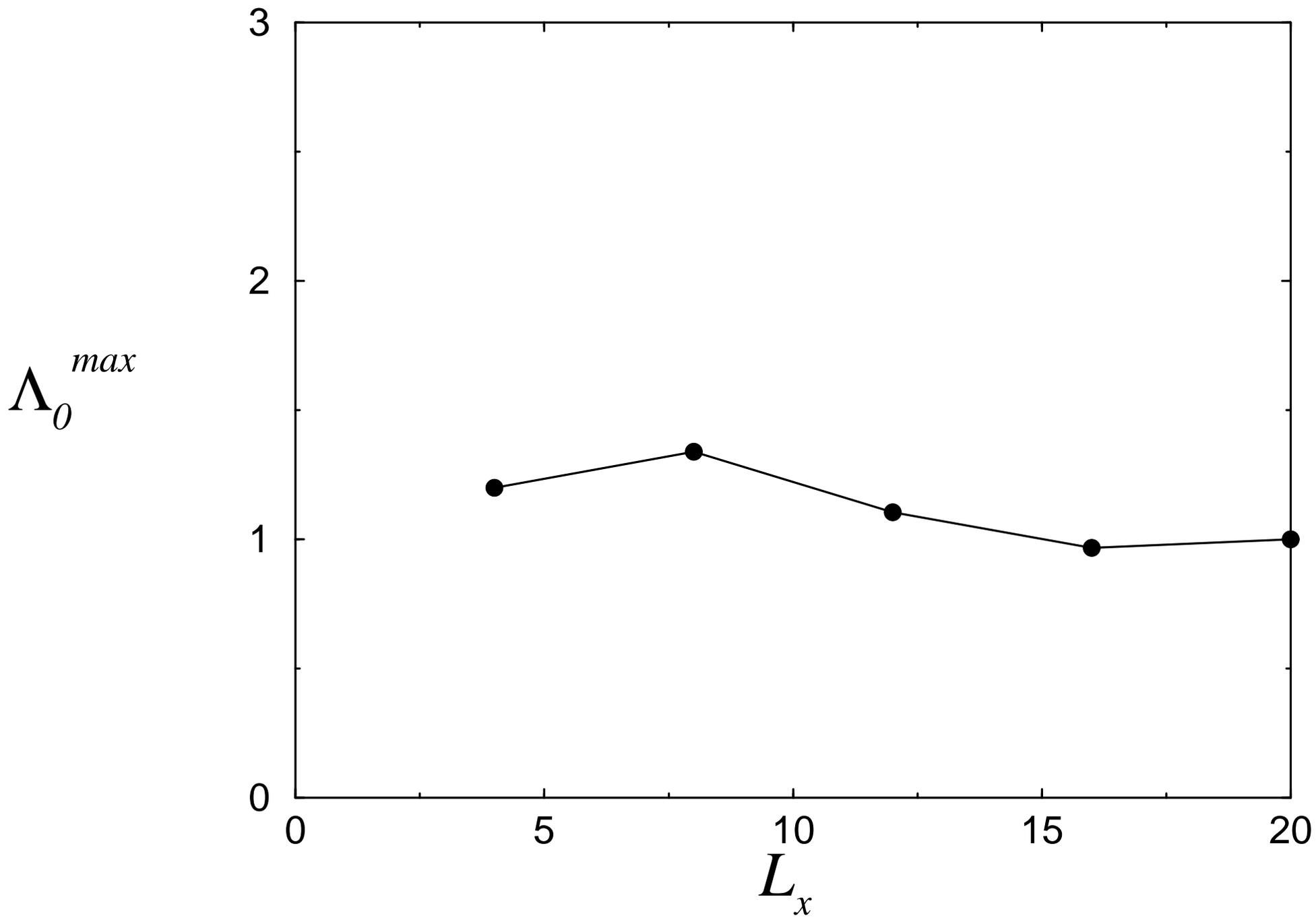
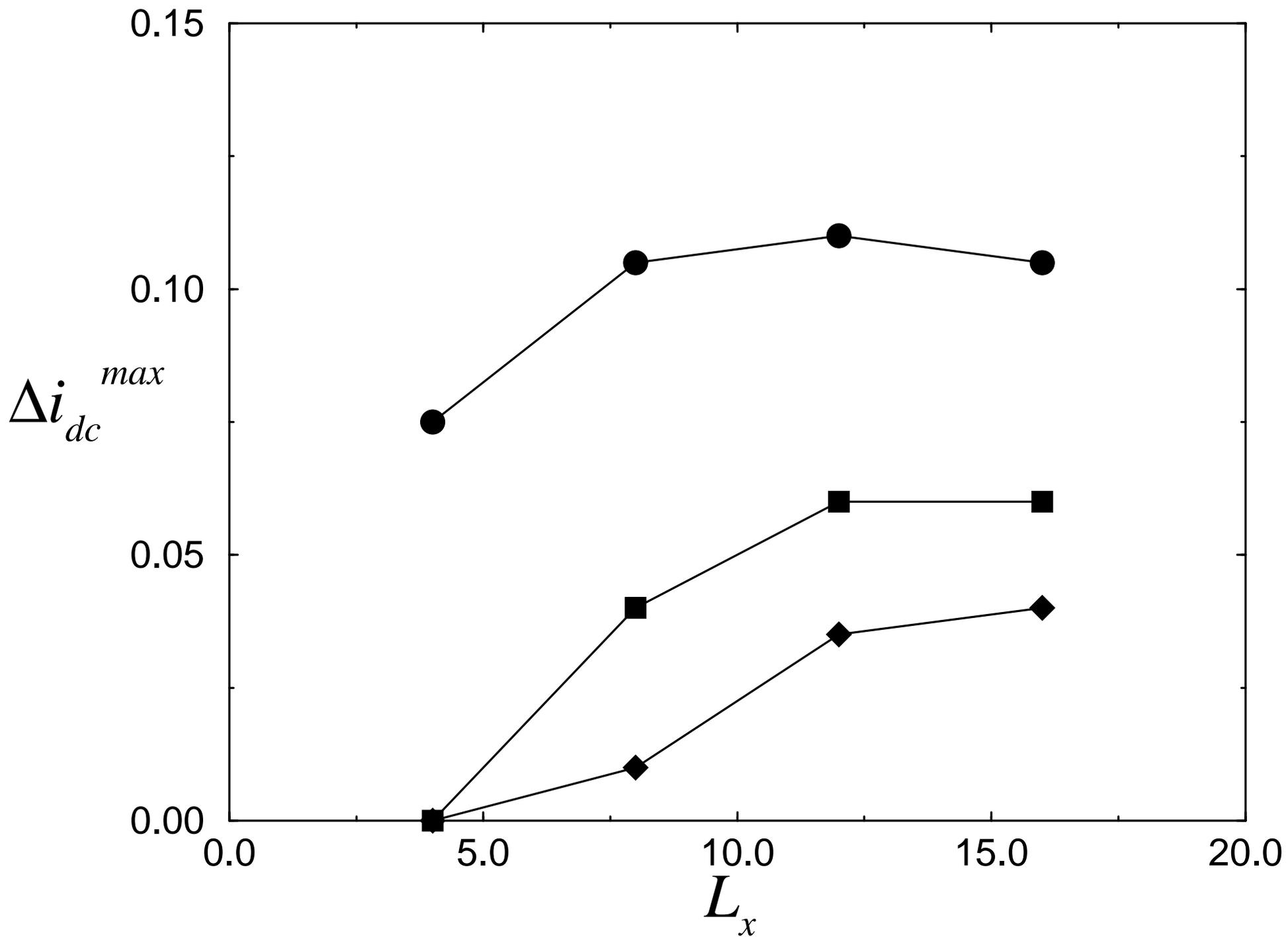
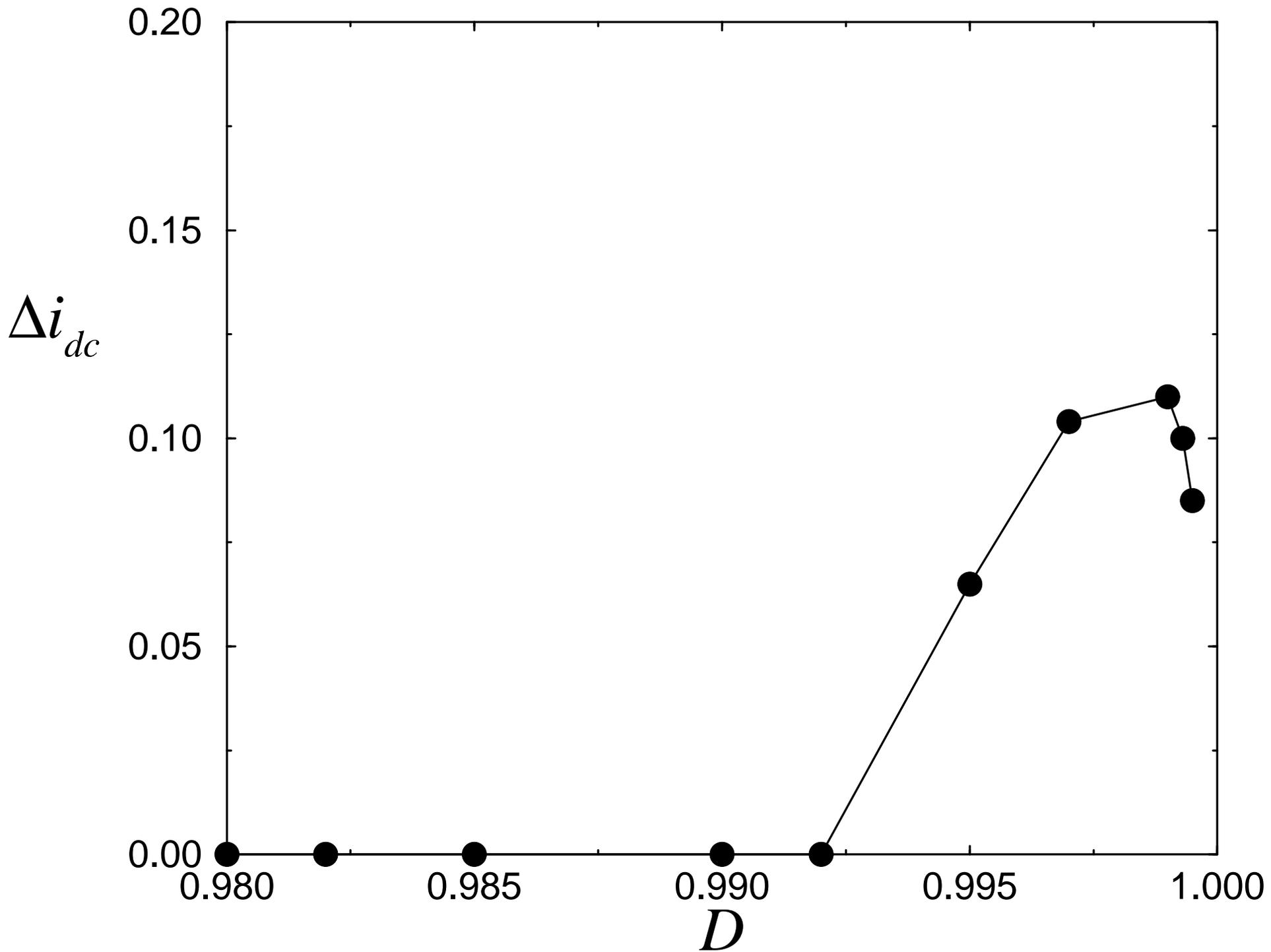

For {\it Model C}, we also investigated the size of the fractional
steps as a function of the disk diameter, $D$. Figure~\ref{i_D_lx16}
shows the dependence of the width of the $\frac{1}{15}$-step in the
$16\times16$ array. The qualitative behavior is very similar to that
of the $8\times8$ array. Quantitatively, the difference lies in the
values at which the steps appear and subsequently disappear. In the
$8\times8$ array, the region of $D$-values for which the steps are
visible is $D \in [0.950, 0.995]$, while in the $16\times16$ array, it
is much smaller than that: $D \in [0.992, 0.9999]$.

\section{Summary and Discussion}

\label{sec:summary}

In this paper we have considered the general response of a
Josephson-junction array that incorporates the capacitance matrix at
different levels of approximation. We have considered three different
models.
{\it Model A} includes only the mutual capacitance between
nearest-neighbor islands. This model has been studied to some extent
before\cite{hagenaars}.
{\it Model B} includes the diagonal terms of $C_{\vec{r},\vec{r}'}$
and had not been studied before. The important element in {\it Model
B} is the {\it exponential} screening present at long distances that
introduces important changes in the current response in the array: to
wit, the new giant capacitive fractional steps (GCFS) that arise in
the I-V's in this model, appearing and disappearing as the screening
length $\Lambda_0$ changes.
{\it Model C} incorporates the electrostatic interaction between all
conducting islands. Since there is no exact solution known for an
exact full capacitive matrix, we considered different model
approximations to the full $C_{\vec{r},\vec{r}'}$. Most of our results
were based on an explicit solution for the capacitance of two disks of
finite size in the plane using a superposition to construct the full
matrix. This approach is similar to the one previously used in the
full inductance-matrix problem\cite{modb}. We also used the
numerically exact {\sc fastcap} algorithm developed at
MIT\cite{fastcap}. The essence and main result from all these
calculations is that the full capacitance decays algebraically rather
than exponentially as in {\it Model B}. We then concluded that the
two-disk approximation contains, qualitatively at least, the essential
properties of the full capacitance matrix of the more accurate
problem.

Our results indicate that for screening lengths $\Lambda_0$ much
larger than the lattice size, the I-V's display the characteristic
giant Shapiro steps of the mutual capacitance model only ({\it Model
A}\cite{hagenaars}). That is, there are giant integer steps, and also
(for certain values of $\beta_c$) the much smaller half-steps that are
triggered by the onset of chaos in a single junction. The half-steps
display the characteristic ACVS. The essential new result from our
calculations in {\it Models B} and {\it C} is that there are new GCFS
that essentially depend on the amount of screening in the model. The
GCFS are not produced by oscillating vortices, as in the
non-full-capacitive problems, but are {\it localized} in restricted
areas of the lattice due to electrostatic screening.

Generally, the size of the GCFS varies inversely with (though not
necessarily proportionally to) the order: that is, lower-order steps
are larger than higher-order ones. These fractional steps in {\it
Model B} are metastable in nature, (that is, history-dependent but
stable once produced) but remain for the longest runs we have carried
out. Once $\Lambda_0\sim 1$, all steps begin to disappear, and by
$\Lambda _0=0.5$ they have completely vanished.

For {\it Model C} when $D=1.0$ (and for large enough screening length,
$\Lambda _0$), we have the under-damped results of {\it Model A}, and
only see integer steps in the I-V's. As $D$ decreases from $1.0$, we
enter a regime in which different rows have different values of
$i_{dc}$ at which they assume the coherent, nonzero voltage state.
The rows closer to the boundaries of the array maintain the critical
current of the under-damped array, while those towards the center
become increasingly difficult to switch on. The interplay between the
switching-on values for different rows determines the relative sizes
of GCFS that we see. There may be no value of $i_{dc}$ for which all
the junctions in the array can oscillate in phase. This corresponds to
both the shrinking of the integer steps, and the appearance of the
fractional steps. As $D$ decreases further, we enter a regime in which
the coherent oscillation is no longer possible even for a single
row. At this point, we have lost {\em all} the steps, fractional and
otherwise.

The question naturally arises: To what extent does this behavior
depend on the model used to construct the capacitance matrix? We
believe this behavior to be independent of the model used, based on
simulations we have carried out using results obtained independently,
and by a completely different method by Whan {\em et al.}\cite{whan}
and the {\sc fastcap} algorithm calculations. The idea now is to check
the predictions of this paper experimentally.

\section*{Acknowledgments}

The authors would like to acknowledge fruitful discussions with Paul
Tiesinga, Thomas Hagenaars and Eugenio Ley-Koo. We thank J. White for
telling us how to obtain {\sc fastcap}. This work was supported in
part by NSF grants DMR-95-21845 and DMR-9821845.


\appendix 
\section*A

In this appendix, we describe how to obtain the capacitance of two
disks given in Eqs.~(\ref{2cap_disk1}) and (\ref{2cap_disk2}) used in
the array calculations. It is best to use bipolar coordinates (see
{\bf Ref}.~\onlinecite{arfken}) to obtain the relevant expression but then
we shall transform to Cartesian coordinates as used in the
calculations\cite{jackson,ley}. These coordinates are related to the
Cartesian coordinates through the equations
\begin{equation}
x= \frac{a\sinh\eta}{(\cosh\eta -\cos\xi)},\qquad
y= \frac{a\sin \xi }{(\cosh\eta -\cos\xi)},\qquad 
z=z.\label{a1}
\end{equation}
The parameter $\eta$ measures the diameter of the disks. $\eta =
-\infty$ represents an infinitesimally small disk at position $(-a,
0)$. As $\eta$ increases, the disk grows becoming infinite in size, at
$\eta = 0$. Increasing $\eta$ further causes the disk to shrink until
it becomes a point at $(a, 0)$, for $\eta = +\infty$. The parameter
$a$ denotes the distance of the points $\eta=\pm\infty$ from the
$y$-axis, $a\coth \eta$ is the distance from the center of the disks
to the $y$-axis, i.e., half of the separation distance between the
centers of the disks. The lines of constant $\eta $ represent the
boundaries of the disks.

The orthogonality of the bipolar coordinates can be established by
evaluating the infinitesimal displacement from Eq.~(\ref{a1}) and
identifying the respective scale factors and unit vectors:
\begin{eqnarray*}
d\vec{r} &=& \hat{e}_x dx + \hat{e}_y dy + \hat{e}_z dz 
= \hat{e}_\xi h_\xi d\xi + \hat{e}_\eta h_\eta d\eta + \hat{e}_z h_z dz,\\
h_\xi &=& h_\eta =\frac {a}{\cosh\eta -\cos\xi},\qquad h_z=1,\\
\hat{e}_\xi&=&\frac{-\hat{e}_x\sinh\eta \sin\xi
+ \hat{e}_y(\cosh\eta \cos\xi -1)}{\cosh\eta -\cos\xi },\\
\hat{e}_\eta&=& \frac{-\hat{e}_x(\cosh\eta\cos\xi -1)
-\hat{e}_y\sinh\eta\sin\xi}{\cosh\eta -\cos\xi}.
\end{eqnarray*}
The electrostatic potential function between the two disks satisfies
the Laplace equation, so from the previous equations we get,
\begin{equation}
\frac {1}{h_\xi h_\eta} \left[ \frac{\partial ^2}{\partial \xi^2}
+ \frac{\partial^2}{\partial\eta^2}\right] \phi(\xi,\eta )=0.\label{3a}
\end{equation}

The general solution to Eq.~(\ref{3a}) can be written as
\begin{eqnarray}
\phi (\xi ,\eta )=\sum_{m=0}^\infty &&\left[ A_m\cos (m\xi )+B_m\sin (m\xi
)\right]\nonumber\\ 
&\times& \left[ C_m\cosh (m\eta )+D_m\sinh (m\eta )\right].\label{4a}
\end{eqnarray}
Taking the disks with centers at $\eta =\eta _1$ and $\eta =\eta _2$,
the boundary conditions are ${\phi (\xi ,\eta =\eta _1)=V_1,}$ and
${\phi (\xi ,\eta =\eta _2)=V_2=0.}$ To get the electrostatic
potential function we only need the case $m=0$:

\begin{equation}
\phi (\xi ,\eta )=\frac{V_2(\eta -\eta_1) +V_1(\eta _2-\eta )}
{\eta_2-\eta_1}. \label{6a}
\end{equation}
The electric field intensity is obtained by taking the negative
gradient of Eq.~(\ref{6a}), and only has $\hat{e}_\eta$-direction
components:

\begin{equation}
\vec{E}=-\vec{\nabla}\phi =-\frac{V_2-V_1}{h_\eta(\eta_2-\eta_1)}
\hat{e}_\eta. \label{7a}
\end{equation}

The electric charge distribution on each of the disks is given by

\begin{eqnarray}
\sigma (\xi ,\eta =\eta_1) & = &\frac{\hat{e}_\eta\cdot\vec{E}}{4\pi }
\Bigg|_{\eta =\eta_1}\nonumber\\
&=&-\frac{(V_2-V_1)(\cosh\eta _1-\cos \xi )}
{4\pi(\eta_2 - \eta_1)a}, \nonumber\\
\sigma (\xi ,\eta =\eta_2) & = & 
-\frac{\hat{e}_\eta \cdot\vec{E}}{4\pi } \Bigg|_{\eta =\eta_2}\nonumber\\
&=&\frac{(V_2-V_1)(\cosh\eta_2-\cos\xi )}{4\pi(\eta_2 - \eta_1)a}. \label{8a}
\end{eqnarray}

The total charges are obtained by integrating Eqs.~(\ref{8a}) over the
respective disks
\begin{eqnarray}
Q_1 &=& \int_0^h\int_0^{2\pi }\sigma (\xi ,\eta =\eta _1) h_\xi d\xi dz\nonumber\\
&=& -\frac{V_2-V_1}{4\pi (\eta_2-\eta_1)}2\pi h
= -\frac{V_2-V_1}{2{(\eta_2-\eta_1)}}h. \label{9a}
\end{eqnarray}

Therefore the capacitance of the bipolar capacitor with height $h$,
follows from Eq.~(\ref{9a}),

\begin{equation}
C=\frac{Q_1}{(V_1-V_2)h}=\frac {1}{2(\eta _2-\eta _1)}. \label{10a}
\end{equation}
Now to use this result in our calculations we need to transform it
appropriately to the lattice structure of the square lattices
considered in this paper. Eq.~(\ref{10a}) assumes that the two disks
are different. In the arrays we have studied the disks have the same
radius and thus $\eta _1=\eta $ and $\eta _2=-\eta _1=-\eta$, so that
the capacitance of the two disks is
\begin{equation}
C=\frac {1}{4\eta}. \label{11a}
\end{equation}
Now going back to Cartesian coordinates, we can express $\eta$ and
$\xi$ as follows:

\begin{eqnarray}
\eta & = &\tanh^{-1}\left[\frac{2ax}{a^2 + x^2 + y^2}\right],\label{eq:eta}\\
\xi & = &\tanh^{-1}\left[\frac{2ax}{a^2 - x^2 - y^2}\right].\label{eq:xi}
\end{eqnarray}
We define $d$ as the ratio of the radius of the disk to the distance
of its center from the origin (or, equivalently, the ratio of the
diameter to the distance between the centers of two adjacent disks):
\begin{equation}
d = \frac{a |\mathrm{csch}\,\eta|}{a\coth\eta} = \frac{1}{\cosh\eta}.
\end{equation}
This provides a way to parameterize the separation between adjacent
disks. Since $\cosh x \equiv (e^x + e^{-x})/2 \geq 1, \,\forall x$,
$d$ lies in the range $[0, 1)$. When $d\lesssim 1$, the disks are
almost touching, and each disk occupies almost all of the lattice unit
cell. We expect this to correspond to {\em Model A}. When $d \simeq
0$, then the disks are separated by a very large distance in relation
to their diameter. This scenario corresponds to weak screening. We
will express the capacitance in terms of this $d$ parameter.

Taking $y=0$, and measuring distances in units of $a$ hereafter,
Eq.~(\ref{eq:eta}) becomes
\begin{equation}\label{eq:eta_x}
\eta = \tanh^{-1}\left[\frac{2x}{1 + x^2}\right].
\end{equation}
The points where the disk intersects with the $x$-axis are given by
$(x - \coth\eta)^2 =\mathrm{csch}^2\eta$ (the equation of a circle of
radius $\mathrm{csch}\eta$, centered at $\coth\eta$). This has
solutions $x = (\cosh\eta\pm 1)/\sinh\eta$, which can be expressed in
terms of $d$:
\begin{equation}\label{eq:x_d}
x = \frac{1\pm d}{\sqrt{1-d^2}}
\end{equation}
(There are two values of $x$ for each $d$---each gives the same
expression for the capacitance, as it should.) Plugging the solutions
in Eq.~(\ref{eq:x_d}) into Eq.~(\ref{eq:eta_x}) gives us:
\begin{eqnarray}\label{eq:eta_d}
\eta &=& \tanh^{-1}\left[ \frac{2(1\pm d)}{\sqrt{1-d^2}}
\frac{1-d^2}{1 -d^2+ (1\pm d)^2}\right]\nonumber\\
& = &\tanh^{-1}\left[\sqrt{1 - d^2}\right].
\end{eqnarray}
Therefore, the capacitance behaves as: 
\begin{equation}
C = \frac{1}{4\tanh^{-1}\left[\sqrt{1-d^2}\right]},
\end{equation} 
where $d$ is the ratio of the disk diameter to the distance between
the centers of two adjacent disks. This is the result used in our
calculations and given in Eqs.~(\ref{2cap_disk1}) and
(\ref{2cap_disk2}).

As a check on this result, we can look at the behavior as
$d\rightarrow 0$. This problem has been solved by
Jackson\cite{jackson} (problem~1.7, p.~51) who gives the capacitance
per unit length for a pair of infinite parallel wires, with a ratio of
radius to separation $d~\ll 1$, as $C = -1/4\ell n\, d$.

By using the expression~\cite{gradstein}:
\[
\tanh^{-1} z = \frac{1}{2}\ell n\,\left[\frac{1 + z}{1 - z}\right],
\]
and expanding binomially the square root of Eq.~(\ref{eq:eta_d}) in
the limit of small $d$, we find that
\begin{equation}
\tanh^{-1}\left[\sqrt{1-d^2}\right] 
\rightarrow \frac{1}{2}\ell n\,\left(\frac{4}{d^2}\right) =
\ell n\,(2/d) \simeq -\ell n\,d.
\end{equation}
This gives us
\begin{equation}
C(d)_{d\rightarrow 0} \simeq \frac{-1}{4\ell n\,d}.
\end{equation}
Thus, we have agreement between our model and a known solution in the
limit of small $d$.



\begin{figure}
\caption{Diagram of the array studied. The symbols $\boxtimes$ denote
the capacitive junctions. Current is injected from the bottom, and
extracted from the top, with periodic boundary conditions along the
direction perpendicular to the current flow.
\label{array_fig}}
\end{figure}
\begin{figure}
\caption{This figure illustrates the two-disk model. $D$ is their
diameter {\em measured in units of the lattice spacing} and $R$ the
separation distance between them. When $D\lesssim 1$, the array is
close-packed, with $D$ almost equal to the lattice spacing. When $D\ll
1$, the array is loosely packed, with the disks widely spaced from
each other, and thus no current can flow between them. Our model is
not valid for this value of $D \ll 1$.\label{packing}}
\end{figure}
\begin{figure}
\caption{Comparison of behavior of  capacitances for different model 
calculations. The lines represent the following models:
{\sc FASTCAP} calculation~(-~-~-); {\it Model C} two-disk
approximation~(--~--~--) for $D=0.99$, and~(--~-~--~-~--) for $D=0.1$;
finite-potential disk embedded in an infinite zero-potential
conducting plane~(---~---~---). The inset shows the conductor plate
array used in the {\sc fastcap} calculations.
The plates are of side $\ell$, with thicknesses ranging
from 1000 to 35000~\AA.
\label{caps_power}}
\end{figure}
\begin{figure} 
\caption{{\it Model C} I-V's for a DC-driven array, as a function of
$D$, for an $8\times8$ array. The solid line was obtained with $D =
0.20$, the dashed one with $D=0.99999$. $\beta_c = 0.50$.
\label{dc_iv}}
\end{figure}
\begin{figure} 
\caption{{\sc fastcap} model I-V's for a DC-driven array, as a
function of the ratio of the diameter $\tilde{D}$ to the lattice
spacing for an $8\times 8$ array studied by Whan {\em et
al.}[19].  The solid line was obtained with
$\tilde{D}=0.050$, the dashed one with $\tilde{D}=0.970$. Parameter
values are $\beta_c = 0.50$.  We use a different diameter value than
in the two-disk case since they are not exactly the same.
\label{FASTCAP_dc}}
\end{figure}
\begin{figure}
\caption{{\it Model B} I-V curves for $40\times40$ array illustrating
the GCFS of decreasing size and visible at $\frac{1}{39}$,
$\frac{2}{39}$, $\frac{3}{39}$, $\frac{4}{39}$(just). The inset shows
the same steps, with the $y$-coordinate multiplied by 39 to show the
precise values of the steps. $\beta_c = 0.50, \nu = 0.10$, $\Lambda_0
= 1.75$. The $\Lambda_0$ value chosen corresponds to one chosen in
previous figures.
\label{lx40_steps}}
\end{figure}
\begin{figure}
\caption{Four I-V's for {\it Model B}, corresponding to
$\Lambda_0= 32, 8, 3, 1.75$, from left to right, and each displaced
for clarity successively by 0.5 units to the right. They show how the
integer $n=1$ step is rapidly destroyed as $\Lambda_0$, is reduced
below the array size ($L_x = 16, L_y = 16$). Just as the integer step
disappears, a fractional step emerges (a central subject of this
paper). Parameter values are $\beta_c = 0.5$, $\nu = 0.1$, and
$i_{\text{ac}} = 1.0$.
\label{stepdisappear}}
\end{figure}
\begin{figure}
\caption{The $n=1$ step width is shown as a function of $\Lambda_0$,
for {\it Model B}. It is clear that once the screening length becomes
less than half the array size ($L_x = L_y = 16$), the step is
destroyed. Values for $\beta_c$, $\nu$, and $i_{\text{ac}}$ are the
same as in Fig.~\ref{stepdisappear}. \label{screen}}
\end{figure}
\begin{figure}
\caption{The width of the fractional $\frac{2}{7}$-step is shown as a
function of $\nu$, in {\it Model B}. Values for $\beta_c$,
$i_{\text{ac}}$ are the same as in Fig.~\ref{stepdisappear} and
$\Lambda_0 = 1.75$. Not shown are values of $\nu \in [0.3, 0.6]$, for
which we have verified that there is no step. \label{width_nu}}
\end{figure}
\begin{figure}
\caption{The widths of the $\frac{1}{7}$-, $\frac{2}{7}$-, and
$\frac{3}{7}$-steps (symbols $\blacktriangle$, $\blacksquare$ and
$\blacklozenge$ respectively) as a function of $\beta_c$, in {\it
Model B}. Parameter values for $\nu$, and $\Lambda_0$ are the same as
in Fig.~\ref{stepdisappear}. We have also looked for (but not found)
other steps but did not find them for $\beta_c \leq
10.0$. \label{width_betac}}
\end{figure}
\begin{figure}
\caption{{\it Model B} widths of the $\frac{1}{7}$- and
$\frac{2}{7}$-steps (symbols $\bullet$ and $\blacktriangle$
respectively) shown as functions of $\Lambda_0$. Values for $\beta_c$
and $\nu$ are the same as in Fig.~\ref{stepdisappear}. We also
have looked for other steps for all values of $\Lambda_0\leq 32$. The
inset shows an I-V showing the hysteresis of the fractional
steps in {\it Model B}. The dashed (solid) line indicates data taken
as the current is increased (decreased), as shown by the arrows. Here
$\beta_c = 0.50, \nu = 0.10, \Lambda_0 = 1.75, \delta i_{\text{dc}} =
0.005$.\label{width_lambda}}
\end{figure}
\begin{figure}
\caption{The width of the $\frac{2}{7}$-step as a function of the
diameter, $D$, in {\it Model C}. See Fig.~\ref{packing} for an
explanation of the symbol $D$. $\beta_c = 0.5$, $\nu = 0.1$. Note that
the step exists only for a small range of $D$-values.
\label{ivsD}}
\end{figure}
\begin{figure}
\caption{The width of the $\frac{2}{7}$-step ($\bullet$) as a function
of $\Lambda_0$, for {\it Model C} with $D=0.994$ (this value gives the
maximum step-width for $\Lambda_0 = 5.66$). The step-width in {\it
Model B} is shown for comparison ($\blacklozenge$). $\beta_c = 0.50$,
$\nu = 0.10$, $L_x = L_y = 8$.
\label{modelC_width_lambda}}
\end{figure}
\begin{figure}
\caption{{\it Model C} I-V's for individual rows of array showing
$\frac{2}{7}$-step. The solid line represents the array average, the
dashed line represents the edge junctions, while the dashed-dotted
line represents interior junctions. Parameter values are $\beta_c =
0.50, \nu = 0.10$.  The inset shows the power spectrum for the
$\frac{1}{7}$-step in {\em Model~B}. Parameter values are $\beta_c =
0.50$, $\nu = 0.10$. The drive frequency and harmonics are clearly
visible, as are other peaks at roughly $\frac{4\nu}{7}$ and
$\frac{5\nu}{7}$. We have verified that these latter are transients.
\label{modelC_edge_on}}
\end{figure}
\begin{figure}
\caption{This figure shows the minimum value of the screening length
$\Lambda_0$, as a function of $L_x$, for which the first step is
visible, in {\it Model B}. The data points are shown as $\bullet$,
fitted to the dashed line by $\log\Lambda_0 \sim \log L_x^{0.914\pm
0.058}$.  Values for $\beta_c$, and $\nu$ are as shown in {\bf
Fig}.~\ref{stepdisappear}.
\label{crit_screen_len}}
\end{figure}
\begin{figure}
\caption{ We show, as a function of $L_x$ in {\em Model B}, the value
of the screening length which maximizes the width of the fractional
steps. $\beta_c$, $i_{ac}$, and $\nu$ are as in {\bf
Fig}.~\ref{stepdisappear}. The line is simply a guide to the eye.
\label{screen_len_max}}
\end{figure}
\begin{figure}
\caption{The maximal width $\Delta i^{\text{max}}_{dc}$ of the
fractional steps in {\it Model B}, is shown as a function of lattice
size. The symbols $\bullet$, $\blacksquare$, $\blacklozenge$ represent
the first through third steps respectively (i.e., $\frac{1}{L_x-1}$,
$\frac{2}{L_x-1}$, $\frac{3}{L_x-1}$ etc.). (Not all steps can be
obtained for all lattice sizes.) This step-width appears to saturate,
though not at the same lattice size for all steps.\label{maxwidth_lx}}
\end{figure}
\end{document}